\documentclass[%
 reprint,
 amsmath,amssymb,
 aps,
]{revtex4-2}

\usepackage{graphicx}
\usepackage{dcolumn}
\usepackage{bm}

\newtheorem{definition}{Definition}[section]

\usepackage{algcompatible, newfloat, caption}
\AtBeginEnvironment{algorithm}{\noindent\hrulefill\par\nobreak\vskip-5pt}
\usepackage{hyperref, cleveref}

\DeclareFloatingEnvironment[
    fileext=loa,
    listname=List of Algorithms,
    name=Algorithm,
    placement=tbhp,
]{algorithm}
\DeclareCaptionFormat{algorithms}{\vskip-15pt\hrulefill\par#1#2#3\vskip-6pt\hrulefill}
\usepackage{tikz}
\usetikzlibrary{calc}

\begin{document}

\preprint{AIP/123-QED}

\title{A Practically Scalable Approach to the Closest Vector Problem\\for Sieving via QAOA with Fixed Angles}

\author{Ben Priestley}\thanks{benjamin.priestley@cs.ox.ac.uk}
\affiliation{Department of Computer Science, University of Oxford}
\affiliation{Quantum Software Lab, School of Informatics, University of Edinburgh}
\author{Petros Wallden}\thanks{petros.wallden@ed.ac.uk}
\affiliation{Quantum Software Lab, School of Informatics, University of Edinburgh}

\date{\today}

\begin{abstract}
The NP-hardness of the closest vector problem (CVP) is an important basis for quantum-secure cryptography, in much the same way that integer factorisation's conjectured hardness is at the foundation of cryptosystems like RSA. Recent work with heuristic quantum algorithms \cite{Yan-2022} indicates the possibility to find close approximations to (constrained) CVP instances that could be incorporated within fast sieving approaches for factorisation. This work explores both the \textit{practicality} and \textit{scalability} of the proposed heuristic approach to explore the potential for a quantum advantage for approximate CVP, without regard for the subsequent factoring claims. We also extend the proposal to include an antecedent ``pre-training'' scheme to find and fix a set of parameters that generalise well to increasingly large lattices, which both optimises the scalability of the algorithm, and permits direct numerical analyses. Our results further indicate a noteworthy quantum speed-up for lattice problems obeying a certain `prime' structure, approaching fifth order advantage for QAOA of fixed depth $p=10$ compared to classical brute-force, motivating renewed discussions about the necessary lattice dimensions for quantum-secure cryptosystems in the near-term.
\end{abstract}

\maketitle

\section{Introduction}
The conjectured hardness of integer (prime) factorisation makes it a central problem to modern information security, becoming the foundation for many widespread public-key cryptosystems, such as RSA \cite{Rivest-1978}. No classical algorithm has yet been found -- nor do we ever expect to find one -- for factoring in polynomial time (see e.g. \citet{Zhang-2024}). 

However, it is well known that these cryptosystems are vulnerable to \textit{quantum} adversaries, following Peter Shor's seminal paper uncovering that fault-tolerant quantum computation allows one to factor composite integers in polynomial-time \cite{Shor-1995}. Since its proposal, we have seen many successful experimental demonstrations of Shor's algorithm on trivial problem instances \cite{Lucero-2012, Lanyon-2007, Lu-2007, Martin-Lopez-2012}.

There have been many new proposals for cryptosystems based on different problems that are believed to be secure against any adversary with a large, fault-tolerant quantum computer (see \citet{Bernstein-2017} for a recent review). Most attention is given to \textit{lattice-based cryptography} \cite{Goldreich-1997, Hoffstein-1998, Hoffstein-2001, Hoffstein-2003, Lyubashevsky-2012, Ducas-2013, Bernstein-2018}. Whilst there has been no shortage of counter-arguments against some of these works \cite{Coppersmith-1997, Nguyen-2006, Ducas-2012, Laarhoven-2015a, Laarhoven-2015b, Becker-2016}, proposals based on lattices offer arguably the most promising route to quantum-secure cryptosystems. This is reinforced by the fact that three of the four standardised postquantum secure cryptosystems by NIST are based on lattices (see section 2.3 of \citet{NIST-report-2022}, or the NIST website \cite{NIST-website-2025}).

\subsection{Quantum-accelerated Sieving on a Lattice}

Classical factoring algorithms often use a method called \textit{sieving} to find smooth relation pairs (sr-pairs; a.k.a. fac-relations). Most famously, the quadratic sieve \cite{Pomerance-1984, Davis-1984} and the general number sieve \cite{Lenstra-1993, Briggs-1998} (and their variants) are the fastest known classical algorithms at time of writing (see \citet{boudot-2022}). In these algorithms, sieving is a major bottleneck; e.g. the factoring record set for RSA-250 in 2020 required a computational cost of some 2,700 core years, roughly 90.7\% of which is accounted for by the sieving procedure \cite{boudot-2022}. It is thus very motivating to find a fast approach to the search for sr-pairs -- \textit{the faster we find them, the faster we factor}.

Claus Peter Schnorr has presented a number of theoretical works for lattice-based sieving methods for integer factorisation \cite{Schnorr-1991, Schnorr-2013, Schnorr-2021}. Recently, \citet{Yan-2022} proposed to accelerate Schnorr's lattice sieving by way of nearest neighbour search via the quantum approximate optimisation algorithm (QAOA) \cite{Farhi-2014}.

The principle claim advertised by these works is that this method reduces the spatial requirements for factoring an integer $N$ to $O(\log N/\log\log N)$ qubits. Many have presented strong evidence that this is a considerable underestimate \cite{Grebnev-2023, aboumrad-2023, Khattar-2023, Ducas-2021}, and relies on outdated assumptions in the underlying theoretical framework of \citet{Schnorr-2021}. Moreover, we note the omission of any time-complexity analysis, and thus a disregard for a consideration of the practical utility of their algorithm.

Besides the contention for the suitability of a lattice-based sieving method like Schnorr's \cite{Vera-2010, Ducas-2021}, the claim in \citet{Yan-2022} implies a potential quantum advantage for the (approximate) solving of a particular form of closest vector problem (CVP), which, if realised, has further implications for the security parameters required in lattice-based cryptosystems. We take a particular interest in the QAOA subroutine used to solve this CVP, and specifically with its practical scalability towards larger problem instances.

Furthermore, we extend the subroutine with an antecedent pre-training step for the parameters of the QAOA's ansatz to obtain fixed angles, which allows us to greatly simplify our analyses (as in \citet{boulebnane-2022}, and \citet{brandao-2018}) and dramatically reduce the computational workload. In parallel with \citet{Prokop-2025}, which performs similar analysis for the shortest vector problem (SVP), this is one of the first applications of fixed angles to a cryptographically significant problem, facilitating novel -- and much neglected -- analysis of time complexity.

\subsection{Our Contributions}

For clarity, we now outline our contributions, primarily extending from the works of \citet{Yan-2022} and Schnorr's collection of works for sieving on a lattice \cite{Schnorr-1991, Schnorr-2013, Schnorr-2021}, and from \citet{boulebnane-2022} and \citet{brandao-2018} for QAOA with fixed angles:

\begin{itemize}
    \item A simple yet robust pre-training scheme that dramatically reduces the computational requirements for refining solutions to the CVP.
    \item Novel time-complexity analysis of a QAOA-based sieving method, using our own implementation derived from \citet{Khattar-2023}. The code for our experiments can be found in \cite{CVP-QAOA-code}. 
    \item An optimal scaling for the time complexity and `generalisability' of the method with respect to lattice dimension by our proposed extension. Interestingly, for fixed depth QAOA (for $p$ up to $10$), we achieve polynomial advantage compared to brute-force, that exceeds significantly the ``usual'' Grover-type quadratic speed-up \footnote{We note that this does not conflict any known results on the asymptotic optimality of Grover, since QAOA is not a ``black-box'' oracle algorithm and uses the structure of the problem (via the problem Hamiltonian) in the way the ansatz is constructed.}.
\end{itemize}

In undertaking the above, we provide an empirical analysis of the prospective security of cryptosystems based on lattice problems with respect to newer variational methods, giving indication for the required security parameters (e.g. lattice dimension) within such schemes. This is important insight due to the rapidly growing interest in, and usage of, variational algorithms driven by their suitability for near-term hardware \cite{Cerezo-2021}.

\subsection{Limitations to Application and Scope}

There are two major limitations that should be taken into account considering the expected impact of our results in practice, and for the scope of their utility.

Firstly, since we adopt the underlying framework from \citet{Yan-2022} for the sake of analysis, we inherent their search space: constant size in each basis-vector (dimension), leading to space-complexity $O(n)$ and search space of the form $2^{O(n)}$. However, for similar works with the shortest vector problem (SVP), it is well-known that a size $O(\log n)$ is required for each basis-vector, and thus a search space of $2^{O(n\log n)}$ is needed to ensure that the true Shortest Vector is within the search space \cite{Albrecht-2023, Joseph-2021}. Consequently, we do not have guarantees that the quality of the solution is close to the true Closest Vector, asymptotically. This is also related with limitations of Schnorr's method and with the ultimate claim of factorisation \cite{Schnorr-2021, Yan-2022} (see appendix \ref{appendix:density}). While we are restricted in this same way, we can still say \textit{something} about solution quality; namely, we can give empirical observation on the degree of improvement over the classical guarantees of Babai's nearest plane algorithm \cite{Babai-1986}.

Secondly, by motivating our CVP via a factorisation problem, we are working with a restricted structure in our lattice. As such, our work can be framed as an analysis of `best-case scenario' lattice problems, wherein assumptions about the discrete gaps between basis vectors can be utilised to design an optimistic algorithm. We will show that exponential effort is required even for this constrained lattice, which can be taken as further evidence against the claims of sublinear factoring, though nonetheless imply a quantum advantage for CVP on certain lattices.

The results of this work are most appropriately interpreted as an understanding of the \textit{practical scalability} of this kind of neighbourhood search on a particularly constrained lattice. These constraints allow us to leverage inherent symmetries of the problem to design a highly effective pre-training scheme. Whether this can be generalised, e.g. to arbitrary CVP structures or for tighter approximations, remains an important open question.

\section{Background and Preliminaries}
\subsection{Prime Factorisation and Sieving} \label{sec:background-factoring-and-sieving}

This section gives an overview of the current state of \textit{classical} integer factorisation. We give essential definitions in factoring and number theory, and reduce the problem of factoring to the problem of finding sr-pairs.

\begin{definition}[Integer factorisation problem]
    Given an odd composite integer $N>2$, find the prime factors $p,q$ (with $p<q$) such that $N=p\cdot q$.
\end{definition}

Let $P_n=\{p_i\}_{i=0,\dots,n}$ denote the $n$-th \textit{prime basis}, where $p_i$ is the $i$-th prime number, and $p_0:=-1$ allows us the capacity to represent negative integers. 

\begin{definition}[Smooth number]
    An integer is called \textit{$p_n$-smooth} if all its factors are in $P_n$. We call $p_n$ the \textit{smooth bound}.
\end{definition}

\begin{definition}[Smooth relation pair] \label{def:sr-pair}
     Moreover, a pair of $p_n$-smooth numbers $(u_j,v_j)$ are called a \textit{$p_n$-smooth relation pair} (sr-pair; a.k.a. \textit{fac-relation} in \citet{Schnorr-2021}) if for $e_{i,j},e'_{i,j}\in\mathbb{N}$, we have that
\begin{equation} \label{eq:sr-pair-criterion}
    u_j=\prod_{i=1}^np_i^{e_{i,j}}\phantom{\int}\text{and}\phantom{\int}u_j-v_jN=\prod_{i=0}^np_i^{e_{i,j}'}\ .
\end{equation}
\end{definition}

The method for factoring that forms the foundations for many of the most efficient classical algorithms goes back to works like \citet{kraitchik-1922} and \citet{Morrison-1975}, and was later developed by \citet{dixon-1981}. Much of this discussion comes from \citet{Schnorr-2021}.

Given $n+1$ sr-pairs $(u_j,v_j)$, and taking the quotient of the terms in eq. (\ref{eq:sr-pair-criterion}), we have for $e_{i,j},e_{i,j}'\in\mathbb{N}$ that 
\begin{equation}
    \frac{u_j-v_jN}{u_j}\equiv\prod_{i=0}^np_i^{e_{i,j}'-e_{i,j}}\equiv1\mod N\ ,
\end{equation}
since $u_j-v_jN\equiv u_j\mod N$ for any $(u_j,v_j)$. Now, any solution $t_1,\dots,t_{n+1}\in\{0,1\}$ of the equations
\begin{equation} \label{eq:solutions-to-factoring}
    \sum_{j=1}^{n+1}t_j(e_{i,j}'-e_{i,j})\equiv0\mod2\ ,
\end{equation}
for $i=0,\dots,n$ solves a difference of squares $X^2-1=(X-1)(X+1)=0\mod N$ by 
\begin{equation}
    X=\prod_{i=0}^np_i^{\frac{1}{2}\sum_{j=1}^{n+1}t_j(e_{i,j}'-e_{i,j})}\mod N\ .
\end{equation}

If $X\neq\pm1\mod N$, then this yields two nontrivial factors $\text{gcd}(X\pm1,N)$ of $N$, where $\text{gcd}(\cdot)$ denotes the greatest common divisor algorithm, which may be efficiently computed using Euclid's algorithm. This idea comes from Fermat's method for factoring.

Solutions to (\ref{eq:solutions-to-factoring}) can be obtained within $O(n^3)$ bit operations since the dimension of the linear system is $O(n)$, and so we are free to neglect this minor part of the workload in factoring $N$. Hence, the bottleneck in factoring is in \textit{finding these $n+1$ sr-pairs}.

\subsection{Lattices and Lattice Problems}

\begin{definition} [Euclidean Lattice]
    A (Euclidean) lattice $\mathcal{L}$ is a discrete additive subgroup of $\mathbb{R}^n$; that is, a subset $\mathcal{L}\subseteq\mathbb{R}^n$ that is closed under addition and subtraction, and wherein there exists some $\varepsilon>0$ such that any two distinct lattice points $\mathbf{x}\neq\mathbf{y}\in\mathcal{L}$ are separated by a distance of a least $\|\mathbf{x}-\mathbf{y}\|\geq\varepsilon$.
\end{definition}

For a basis matrix $B=[\mathbf{b}_1,\dots,\mathbf{b}_m]\in\mathbb{R}^{n\times m}$ consisting of $m$ linearly independent column vectors in $\mathbb{R}^n$, the lattice $\mathcal{L}(B)=\{B\mathbf{x}:\mathbf{x}\in\mathbb{Z}^m\}$ is generated by all integer linear combinations of $\mathbf{b}_1,\dots\mathbf{b}_m$. Intuitively, a lattice is simply a regular ordering of points.

Following these semantics, the \textit{dimension} of $\mathcal{L}$ is $n$, and its \textit{rank} is $m$. When $n=m$, the lattice $\mathcal{L}$ (and the matrix $B$) are called \textit{full rank}.

\subsubsection{Properties of lattices}

There are a couple of interesting properties to notice: lattices are (1) dense; and (2) hard to approximate. For a given region, there may be many lattice points, which can make finding specific points difficult. And, given an arbitrary point, finding nearby points is again difficult.

\begin{definition} [Successive minima]
    The successive minima for a lattice $\mathcal{L}\subseteq\mathbb{R}^n$ are the positive values $\lambda_1(\mathcal{L})\leq\dots\leq\lambda_n(\mathcal{L})$, where $\lambda_k(\mathcal{L})$ is the smallest radius of a zero-centred ball containing $k$ linearly independent vectors of $\mathcal{L}$.
\end{definition}

Thus, $\lambda_1=\lambda_1(\mathcal{L})$ is the shortest nonzero vector in $\mathcal{L}$.

\begin{definition} [Hermite constant] \label{def:hermite}
    The minimal $\gamma$ satisfying $\lambda_1^2\leq\gamma(\det\mathcal{L})^{2/n}$ for all lattices of dimension $n$ is called the Hermite constant $\gamma_n$, where $\det\mathcal{L}=(\det B^TB)^{1/2}$ is the determinant of $\mathcal{L}$.
\end{definition}

\subsubsection{Problems on lattices}

\begin{definition} [Shortest Vector Problem] \label{def:svp}
    Given a basis $B$ for a lattice $\mathcal{L}(B)$, find a vector $\mathbf{v}\neq\mathbf{0}\in\mathcal{L}$ such that $\|\mathbf{v}\|=\lambda_1(\mathcal{L})$, where $\|\mathbf{v}\|=(\mathbf{v}^T\mathbf{v})^{1/2}$.
\end{definition}

Often it suffices to be only `close' to the true shortest vector $\lambda_1$. In these cases, we refer instead to an \textit{approximate shortest vector problem ($\alpha$-SVP)} for which the condition in definition \ref{def:svp} is amended as $\|\mathbf{v}\|\leq\alpha\cdot\lambda_1(\mathcal{L})$ for approximation factor $\alpha\geq1$.

The exact value of $\lambda_1$ can, in itself, be hard to obtain due to the inherent hardness of the SVP (and of $\alpha$-SVP). It may then be preferable to instead define the problem according to a (relatively) easily computable value relating to $\lambda_1$. For example, we can again amend the condition in definition \ref{def:svp} to $\|\mathbf{v}\|\leq r\cdot(\det\mathcal{L})^{1/n}$ to obtain the \textit{$r$-Hermite shortest vector problem ($r$-Hermite SVP)}. This allows us to check solutions far more easily and thus efficiently, although we lose accuracy in the comparison with the true shortest vector. 

\begin{definition} [Closest Vector Problem] \label{def:cvp}
    Given a basis $B$ for a lattice $\mathcal{L}(B)$, and a target vector $\mathbf{t}\in\text{Span}(B)$, find a vector $\mathbf{v}\in\mathcal{L}$ such that the distance $\|\mathbf{v}-\mathbf{t}\|$ is minimised; i.e. that $\|\mathbf{v}-\mathbf{t}\|=\text{dist}(\mathcal{L},\mathbf{t})$.
\end{definition}

\begin{figure}[ht]
  \centering
  \begin{tikzpicture}
    \coordinate (Origin)   at (0,0);
    \coordinate (Target)   at (3.3,4.1);
    \coordinate (XAxisMin) at (-5,0);
    \coordinate (XAxisMax) at (20,0);
    \coordinate (YAxisMin) at (0,-5);
    \coordinate (YAxisMax) at (0,20);

    \clip (-2,-2) rectangle (6cm,6cm);
    \pgftransformcm{1}{.7}{.4}{1}{\pgfpoint{0cm}{0cm}}
    
    \coordinate (Bone) at (-2,4);
    \coordinate (Btwo) at (6,-2);
    
    \draw[style=help lines,dashed] (-10,-10) grid[step=2cm] (10,10);
    \foreach \x in {-10,-9,...,10}{
      \foreach \y in {-10,-9,...,10}{
        \node[draw,circle,inner sep=1pt,fill] at (2*\x,2*\y) {};
      }
    }
    
    \draw [thick,-latex,black] (Origin) -- (Bone) node [above left] {$\mathbf{b}_1$};
    \draw [thick,-latex,black] (Origin) -- (Btwo) node [below right] {$\mathbf{b}_2$};
    \draw [thick,-latex,blue] (Origin) -- (2, 2) node [below right] {$\mathbf{v}$};
    \draw [thick,-latex,red] (Origin) -- (2, -2) node [below right] {$\lambda_1(\mathcal{L})$};
    
    \node[draw,circle,inner sep=2pt, fill, black] at (Origin) {};
    \node[draw,circle,inner sep=2pt, fill, blue] at (Target) {};
    \draw [thin,-latex,blue] (Target) node [above right] {$\mathbf{t}$};

  \end{tikzpicture}
  \caption{Visualising a two-dimensional, full rank lattice $\mathcal{L}$ generated by the basis $B=[\mathbf{b_1},\mathbf{b_2}]$. Shown in red is the shortest vector $\lambda_1(\mathcal{L})$ solving the SVP for this lattice. Shown in blue is a CVP with target $\mathbf{t}$, solved by the lattice vector $\mathbf{v}$.}
  \label{fig:lattice-problems}
\end{figure}
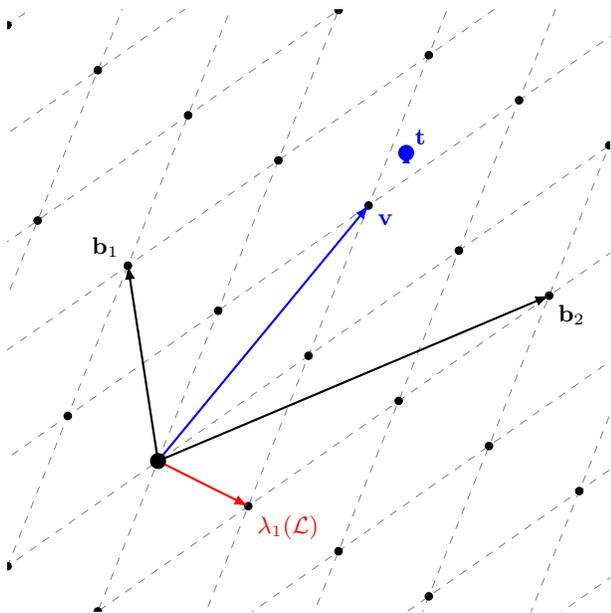

Again, there exists an \textit{approximate closest vector problem ($\alpha$-CVP)} to weaken the condition of closeness in definition \ref{def:cvp}, and the \textit{$r$-approximate closest vector problem ($r$-$Abs$CVP)} weakens the condition further to bring this distance to within $r$, which could, for example, be computed according to $\det\mathcal{L}$ in place of $\text{dist}(\mathcal{L},\mathbf{t})$.

All of these problems are understood to be difficult -- especially in higher dimensions -- to such a degree that even a quantum advantage may be insufficient to yield a polynomial-time solution. Specifically, the decision variant of SVP (GapSVP) is conjectured to be NP-hard \cite{Bennett-2023}, and solving either of SVP or GapSVP within a polynomial factor requires superpolynomial time with quantum computation \cite{Regev-2005}. In general, CVP is thought to be even harder \cite{bennett-2017, micciancio-2001, micciancio-2002}, and is known to be NP-hard \cite{micciancio-2002}.

For the interested reader, \citet{micciancio-2002} remains a prominent text in the complexity of lattice problems for applications in cryptography.

\subsection{Quantum Approximate Optimisation Algorithm} \label{sec:background-QAOA}

A good introduction to variational quantum algorithms in general is offered by \citet{Cerezo-2021}. In this work, we will focus exclusively on perhaps the most widely studied variational algorithm: \textit{quantum approximate optimisation algorithm (QAOA)} due to \citet{Farhi-2014}. We are especially interested in the possibility of pre-training a fixed set of angles (see section \ref{sec:background-fixed-angles}), for which QAOA is an ideal candidate \cite{brandao-2018}.

QAOA was originally inspired by the quantum adiabatic algorithm \cite{Farhi-2000, Farhi-2001}. Adiabatic evolution is replaced with several rounds of propagation between a \textit{problem Hamiltonian} $\hat{H}_C$, which encodes the solution to an optimisation problem $C(z)$ over binary variables $z=z_1,z_2,\dots$ within its ground state, defining the unitary operator
\begin{equation}
    U(\gamma,\hat{H}_C)=e^{-i\gamma \hat{H}_C}\ ,
\end{equation}
and \textit{mixer Hamiltonian} $B$, whose ground-state is known, defining the unitary operator
\begin{equation}
    U(\beta,B)=e^{-i\pi\beta B}=\prod_je^{i\pi\beta \sigma^x_j/2}\ ,
\end{equation} 
parameterised by angles $0\leq\gamma\leq2\pi$ and $0\leq\beta\leq\pi$ respectively, where $\sigma^x_j$ denotes a Pauli-$X$ operator applied on the $j$-th qubit.

We open a uniform superposition over computational basis states to yield an initial state $|s\rangle=\frac{1}{\sqrt{2}}\sum_z|z\rangle$. Then, for some integer number of ansatz layers $p>0$, we define the angle-dependent state
\begin{equation}
    \begin{split}
        |\gamma,\beta\rangle=&U(\beta_p,B)U(\gamma_p,\hat{H}_C)\cdots\\
        &\cdots U(\beta_1,B)U(\gamma_1,\hat{H}_C)|s\rangle\ ,
    \end{split}
\end{equation}
parameterised by angles $\gamma\equiv\gamma_1\dots\gamma_p$ and $\beta\equiv\beta_1\dots\beta_p$.

It is shown in \citet{Farhi-2014} that
\begin{equation}
    \lim_{p\to\infty}\min_{\gamma,\beta}\langle\gamma,\beta|\hat{H}_C|\gamma,\beta\rangle=\min_zC(z)\ ,
\end{equation}
which implies that the optimisation problem can be solved with enough repetitions, if only a good set of angles can be selected. Pessimistically, we can use an outer optimisation loop on a classical computer to search for a good set within the optimisation landscape of $C(z)$ \cite{Cerezo-2021}. Optimistically, there may be precedence to neglect much or all of this optimisation by \textit{fixing these angles}.

\subsection{Fixed Angles for QAOA} \label{sec:background-fixed-angles}

Interesting remarks have been made about the apparent independence between the optimisation landscape for the objective function (of some e.g. combinatorial search problem) and the specific problem instance \cite{Zhou-2020, Bravyi-2019, brandao-2018}. 

In particular, \citet{brandao-2018} fix the $\gamma$ and $\beta$ parameters to show that, when instances are generated by some reasonable distribution, the objective function is nearly independent of the chosen instance. It is suggested that this could be leveraged to find a good set of parameters for some given instance (possibly at great computational expense), which can then be utilised at no further expense for any subsequent (typical) instance. Indeed, removing the need to search for parameters in each instance allows us to neglect the outer optimisation loop once a set has been found, and hence the amortised cost tends to zero inversely with the number of instances being solved \cite{brandao-2018}.

More recently, \citet{boulebnane-2022} have given numerical results that speak to the validity of a ``fixed angle'' scheme for eventually obtaining a quantum advantage with QAOA for random $k$-SAT. They find a significant improvement over Grover's algorithm \cite{Grover-1996} with greatly reduced quantum circuit depth.

In all of these works, the method for obtaining a fixed set of angles is either to take a random (usually small) problem instance to train (e.g. in \citet{brandao-2018}), or the angles may be randomised within some sensible interval. Preliminary investigation with such a scheme led to underwhelming results for this work (see appendix \ref{appendix:alternative-training}), so we seek a more robust pre-training scheme that will find a good set of angles more deliberately. 

\section{Reducing Sieving to a CVP on the Prime Lattice}
The problem of finding small integers whose \textit{product} is close to $N$ is translated into the equivalent problem of finding logarithms of small numbers whose \textit{sum} is close to $\log N$. In doing this, we have given ourselves a combinatorial optimisation problem which may be directly expressed as a linear system of lattice vectors.

Of course, the lattice vectors in question must exhibit the properties of the logarithms of primes so that a linear combination represents an sr-pair. \citet{Schnorr-2021} suggests to construct the so-called \textit{prime lattice} whose basis is defined according to the corresponding factor basis. Some additional randomness is baked in to bring about unique lattice problems that (hopefully) have unique solutions. See appendix \ref{appendix:density} for details concerning Schnorr's method.

Concretely, define the prime lattice $\mathcal{L}(B_{n,c})$ by 
\begin{equation}
    B_{n,c}=
    \begin{pmatrix}
        f(1) & 0 & \cdots & 0 \\
        0 & f(2) & \cdots & 0 \\
        \vdots & \vdots & \ddots & \vdots \\
        0 & 0 & \cdots & f(n) \\
        N^c\ln p_1 & N^c\ln p_2 & \cdots & N^c\ln p_n
    \end{pmatrix}\ ,
\end{equation}
with $c$ the \textit{precision parameter}, and where the $f(i)$ elements correspond to elements from a random permutation of $\{\lceil1/2\rceil,\lceil2/2\rceil,\dots,\lceil n/2\rceil\}$. Hence, the number of lattices arising from a factor basis $P_n$ scales as $O(n!)$.

The target for our CVP on this lattice is defined directly from the composite integer $N$ to be factored:
\begin{equation}
    \mathbf{t}=
    \begin{pmatrix}
        0 & \stackrel{n}{\cdots} & 0 & N^c\ln N
    \end{pmatrix}^\top\ .
\end{equation}

\subsection{Suitability of Approximation}

A critical consideration of the method due to \citet{Schnorr-2021} is that an approximate solution to this CVP is sufficient. Theoretical results \cite{Schnorr-1991, Schnorr-2013, Schnorr-2021} leave the door open to the existence of an approximate solution that improves on polynomial-time classical approximations, but which does not succumb to the NP-hardness of exactly solving the full CVP \cite{bennett-2017, micciancio-2001, micciancio-2002}. We draw from excellent discussion in \citet{aboumrad-2023}, extended from \citet{Schnorr-2021}, to detail the rationale behind this possibility.

Suppose we obtain a set of coefficients $e_j$ for the linear combination of basis vectors of the prime lattice $\mathcal{L}(B_{n,c})$ that approximately solve the CVP. That is, suppose we have found $e_j$ such that
\begin{equation} \label{eq:form-of-solution-to-CVP}
    \varepsilon:=\bigg|\sum_{j=1}^ne_j\ln p_n-\ln N\bigg|\approx0\ .
\end{equation}

If we set that $u=\prod_{e_j\geq0}p_j^{e_j}$ and $v=\prod_{e_j>0}p_j^{-e_j}$, then we obtain $\ln|(\frac{u}{vN})|=\varepsilon$. By Taylor's theorem, $u-vN=vN(e^\varepsilon-1)\approx\varepsilon vN$. \citet{Schnorr-2021}'s argument is that since $\varepsilon$ is small, $u-vN$ is also small and thus \textit{likely} to be $p_n$-smooth. By definition \ref{def:sr-pair}, $(u,v)$ is an sr-pair.

If we are content to run multiple problem instances, then we need only find $e_j$ coefficients for eq. (\ref{eq:form-of-solution-to-CVP}) that reduce $\varepsilon$ to be small enough such that the probability that they correspond to a useful sr-pair is `good enough'. 

There are strong doubts concerning the validity of this method at scale \cite{Grebnev-2023, Khattar-2023, aboumrad-2023, Ducas-2021, Vera-2010}, and in particular that this probability decays exponentially such that sr-pairs cannot be found by refinement alone (or, at least, not quickly enough to care). More detail is given to these concerns in appendix \ref{appendix:density}. 

This work does not look to affirm nor deny Schnorr's algorithm as a method for factoring; our interest lies exclusively with the efficiency with which this CVP can be `refined' by a variational approach, and thus what we might say about the quantum advantage one might find for problems on the prime lattice.

\subsection{Hyperparameters}

So far, we can see that the hyperparameter $n$ has the role of dictating all of: (1) the dimension of the lattice; (2) the size of the factor basis, and hence the number of integers that may be considered $p_n$-smooth; and (3) the number of unique prime lattices that can be constructed in our sieving procedure. 

Increasing $n$ comes with a dimension-complexity trade-off: with a larger smooth bound $p_n$, we can more easily find smooth numbers, but we will require more of them (recall we require $n+1$ sr-pairs in section \ref{sec:background-factoring-and-sieving}). The exact correlation between $n$ and time complexity is not well understood in this context, and much of our work here is dedicated to uncovering this relationship empirically.

The $N^c$ term in the formulation of $B_{n,c}$ gives parameterisable precision to the lattice. This is conjectured to be positively correlated with the probability of finding solutions to problems within this lattice, though some have voiced concerns that this is unsubstantiated \cite{Khattar-2023, Ducas-2021}. As for our later analyses, we will fix $c$ and, similar to \citet{Yan-2022} and \citet{Khattar-2023}, we exchange $N^c$ for $10^c$ to give a consistent basis between instances. This will improve the effectiveness of pre-training.

\begin{figure*}
    \centering
    \textbf{Refinement Performance with Increasing Lattice Dimension (by Depth $p$)}\par\medskip
    \includegraphics[width=.7\textwidth]{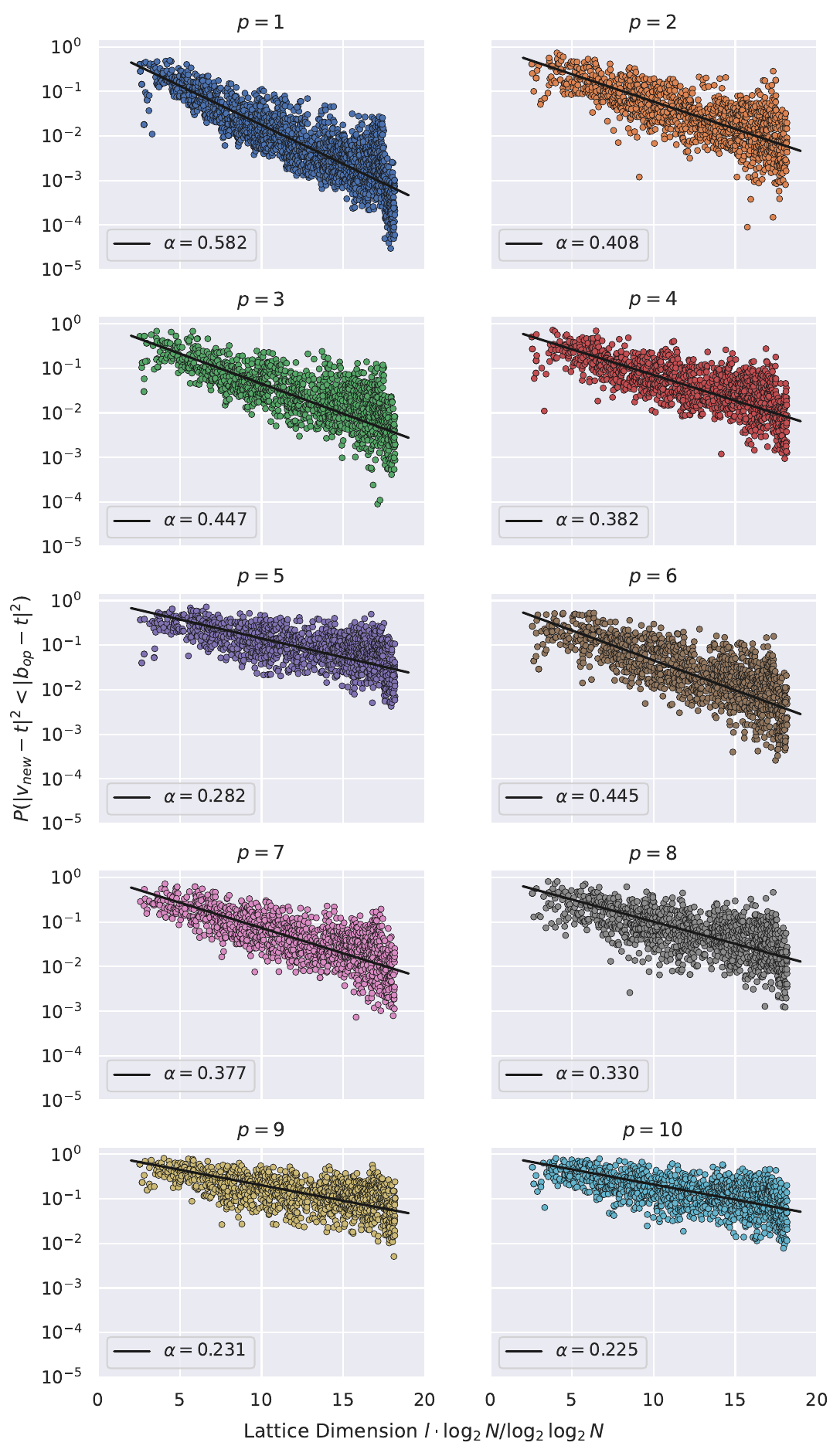}
    \caption{The probability to refine the classical solution (for cases in which a refinement exists) by exact lattice dimension for different depths $p$ of QAOA circuits. Each plot is equipped with a best-fit curve of the form $1/2^{\alpha n}$, and $\alpha$ is shown.}
    \label{fig:performance-by-n}
\end{figure*}

\section{A Method for CVP Refinement} \label{sec:method}
\subsection{Polynomial-time Classical Approximation} \label{sec:classical-approximation}

We begin by considering a polynomial-time procedure that serves up an approximate solution to the CVP. This procedure consists of a lattice reduction step using the LLL-reduction algorithm \cite{Lenstra-1982} (described in appendix \ref{appendix:lattice-reduction}) followed by Babai's nearest plane algorithm \cite{Babai-1986}, which brings us to within a factor $2(2/\sqrt{3})^n$ of $\text{dist}(\mathcal{L},\mathbf{t})$. The following is a high-level summary of Babai's algortihm.

Having produced an LLL-reduced basis $D=[\mathbf{d}_1,\dots,\mathbf{d}_n]$ for $B_{n,c}$, we perform Gram-Schmidt orthogonalisation (without column normalisation) to yield $\tilde{D}=[\tilde{\mathbf{d}}_1,\dots,\tilde{\mathbf{d}}_n]$. The nearest plane algorithm then comprises a series of projections of the target vector $\mathbf{t}$ onto $\text{Span}(\tilde{D})$ followed by a rounding of the coefficients to snap onto a nearby lattice point. Concretely, begin by setting $\mathbf{b}_{op}\leftarrow\mathbf{t}$, then for each $j$ in $n,\dots,1$ compute the Gram-Schmidt coefficient $\mu_j=\langle\mathbf{t},\tilde{\mathbf{d}}_j\rangle/\langle\tilde{\mathbf{d}}_j,\tilde{\mathbf{d}}_j\rangle$, round to the nearest integer $c_j=\lceil\mu_j\rfloor$, and update $\mathbf{b}_{op}\leftarrow\mathbf{b}_{op}-c_j\mathbf{d}_j$.

Intuitively, when considering index $j$, we are taking the $j$-dimensional subspace $\text{Span}\{\tilde{\mathbf{d}}_1,\dots,\tilde{\mathbf{d}}_j\}$ and finding the integer $c_j$ that minimises the distance from $c_j\mathbf{b}_j+\text{Span}\{\tilde{\mathbf{d}}_1,\dots,\tilde{\mathbf{d}}_{j-1}\}$ to $\mathbf{t}$ \cite{aboumrad-2023}; in each step, we find the \textit{nearest (hyper)plane}, giving the algorithm its name.

\subsection{Refinement as a Minimum-eigenstate Optimisation Problem} \label{sec:minimum-eigenstate-problem}

Suppose we have found an approximate solution $\mathbf{b}_{op}=\sum_{j=1}^nc_j\mathbf{d}_j$, where again the $c_j=\lceil\mu_j\rfloor$ are the rounded Gram-Schmidt coefficients. Our goal is to `refine' this solution efficiently.

The potential for a quantum advantage falls out of the rounding operation $\lceil\cdot\rfloor$; classically, this operation considers rounding in each direction one after another (i.e. $\lceil\cdot\rceil$ and $\lfloor\cdot\rfloor$), but quanutmly they can be considered simultaneously \cite{Yan-2022}. Doing this by classical means increases the number of operations exponentially, making it infeasible. \citet{Yan-2022} propose to leverage the effect of superposition to simultaneously encode the two values obtained by each rounding operation within qubits.

To this effect, we will search in the unit neighbourhood centred on $\mathbf{b}_{op}$ to capture all possible rounding arrangements for the coefficients, looking for that which is closest to the target vector $\mathbf{t}$. Concretely, any new vector $\mathbf{v}_{new}$ is obtained by randomly floating $x_j\in\{0,\pm1\}$ on the $c_j$ coefficients, where this $\pm$ comes from the rounding already applied in Babai's nearest plane algorithm;
\begin{equation}
    \mathbf{v}_{new}=\sum_{j=1}^n(c_j+x_j)\mathbf{d}_{j}=\mathbf{b}_{op}+\sum_{j=1}^nx_j\mathbf{d}_j\ .
\end{equation}

To clean up the complication of hard-coded $\pm$s, we can set $x_j=\text{Sign}(\mu_j-c_j)\cdot z_j=\kappa_jz_j$, where we now have binary variables $z_j$. Correspondingly,
\begin{equation}
    \mathbf{v}_{new}=\mathbf{b}_{op}+\sum_{j=1}^n\kappa_jz_j\mathbf{d}_j\ .
\end{equation}

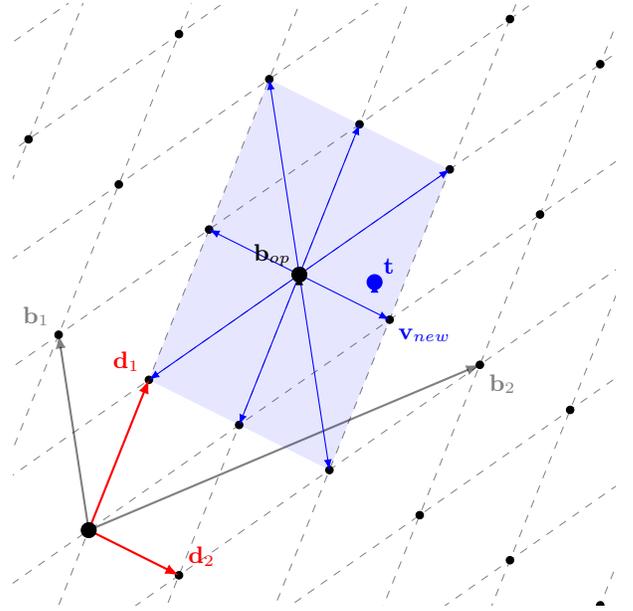
\begin{figure}[ht]
  \centering
  \begin{tikzpicture}
    \coordinate (Origin)   at (0, 0);
    \coordinate (Target)   at (3.8,3.3);
    \coordinate (XAxisMin) at (-5,0);
    \coordinate (XAxisMax) at (20,0);
    \coordinate (YAxisMin) at (0,-5);
    \coordinate (YAxisMax) at (0,20);

    \clip (-1,-1) rectangle (7cm, 7cm);
    \pgftransformcm{1}{.7}{.4}{1}{\pgfpoint{0cm}{0cm}}
    
    \coordinate (Bone) at (-2,4);
    \coordinate (Btwo) at (6,-2);
    \coordinate (Done) at (2,-2);
    \coordinate (Dtwo) at (0, 2);
    
    \draw[style=help lines,dashed] (-10,-10) grid[step=2cm] (10,10);
    \foreach \x in {-10,-9,...,10}{
      \foreach \y in {-10,-9,...,10}{
        \node[draw,circle,inner sep=1pt,fill] at (2*\x,2*\y) {};
      }
    }
    
    \draw [thick,-latex,black, opacity=.5] (Origin) -- (Bone) node [above left] {$\mathbf{b}_1$};
    \draw [thick,-latex,black, opacity=.5] (Origin) -- (Btwo) node [below right] {$\mathbf{b}_2$};
    \draw [thick,-latex,red] (Origin) -- (Done) node [above right] {$\mathbf{d}_2$};
    \draw [thick,-latex,red] (Origin) -- (Dtwo) node [above left] {$\mathbf{d}_1$};

    \node[draw,circle,inner sep=2pt, fill, black] at (Origin) {};

    \draw[thin, -latex, blue] (2, 2) -- (2, 0);
    \draw[thin, -latex, blue] (2, 2) -- (2, 4);
    \draw[thin, -latex, blue] (2, 2) -- (0, 2);
    \draw[thin, -latex, blue] (2, 2) -- (4, -2);
    \draw[thin, -latex, blue] (2, 2) -- (0, 4);
    \draw[thin, -latex, blue] (2, 2) -- (4, 2);
    \draw[thin, -latex, blue] (2, 2) -- (4, 0) node [below right] {$\mathbf{v}_{new}$};
    \draw[thin, -latex, blue] (2, 2) -- (0, 6);
    \filldraw[fill=blue, fill opacity=0.1, draw=none] (0, 2) -- (0, 6) -- (4, 2) -- (4, -2) -- cycle;

    \node[draw,circle,inner sep=2pt, fill, blue] at (Target) {};
    \draw [thin,-latex,blue] (Target) node [above right] {$\mathbf{t}$};
    \node[draw,circle,inner sep=2pt, fill, black] at (2,2) {};
    \draw [thin,-latex,black] (2,2) node [above left] {$\mathbf{b}_{op}$};

  \end{tikzpicture}
  \caption{Visualisation for the nearest neighbour search centred on the approximate solution $\mathbf{b}_{op}$. The LLL-reduced basis $D=[\mathbf{d}_1,\mathbf{d}_2]$ has been computed and used to define a local neighbourhood (shaded blue region) that aims to encompass lattice points that may be closer to the target $\mathbf{t}$. In this case, $\mathbf{v}_{new}$ is shown with floated values $x_1=-1$ and $x_2=1$, and represents a refinement to the approximate solution.}
  \label{fig:nearest-neighbour-search-lattice}
\end{figure}

The cost function in the quadratic unconstrained binary optimisation (QUBO) problem is constructed according to the Euclidean distance between the new vector $\mathbf{v}_{new}$, defined by the bit-string $z_1,\dots,z_n$, and $\mathbf{t}$:
\begin{equation} \label{eq:cost-function}
    C(z_1,\dots,z_n):=\bigg\|\mathbf{t}-\mathbf{b}_{op}-\sum_{j=1}^n\kappa_jz_j\mathbf{d}_j\bigg\|^2\ .
\end{equation}

Any QUBO problem of the form $\sum_ih_iz_i+\sum_{i<j}J_iz_iz_j$ can be expressed by an Ising Hamiltonian -- an energy observable over a system of spin-$1/2$ particles \cite{Grebnev-2023} -- of the form $-\sum_ih_i\sigma_i^z-\sum_{i<j}J_i\sigma_i^2\sigma_j^z$, where $\sigma_i^2$ is the Pauli-$Z$ operator applied on the $i$-th qubit \cite{lucas-2014}. For our cost in eq. (\ref{eq:cost-function}), the Hamiltonian $\hat{H}_C$ can be obtained by directly mapping the binary variables $z_j$ to the Pauli-$Z$ terms $\hat{z}_j=(I-\sigma_j^z)/2$, giving
\begin{equation}
    \begin{split}
        \hat{H}_C=&\bigg\|\mathbf{t}-\mathbf{b}_{op}-\sum_{j=1}^n\kappa_j\hat{z}_j\mathbf{d}_j\bigg\|^2\\
        =&\sum^{n+1}_{i=1}\bigg|t_i-b_{op}^i-\sum_{j=1}^n\kappa_j\hat{z}_jd_{j,i}\bigg|^2\ ,
    \end{split}
\end{equation}
with any negation of terms absorbed into the $\kappa_j$ terms to keep the formulation clean and closer in notation to that of \citet{Yan-2022}.

The energy states (eigenstates) of $\hat{H}_C$ correspond then to the vectors in the unit neighbourhood centred on $\mathbf{b}_{op}$ (including $\mathbf{b}_{op}$ itself), with the corresponding energies (eigenvalues) given by their distance to $\mathbf{t}$. As such, we have formulated a minimum eigenstate problem that may be solved by QAOA (see section \ref{sec:background-QAOA}).

The number of qubits required to refine the approximation is linear $O(n)$ in the dimension of the lattice. Whether this is sufficient to yield a significant enough improvement to lead to efficient sieving is another question, and one that we will briefly explore in section \ref{sec:experiments}. However, it is far more common to search with $O(n\log n)$ resources, and criticism of \citet{Yan-2022} indicate that the linear search space is not enough \cite{Grebnev-2023, Khattar-2023, aboumrad-2023} (further discussion is given in appendix \ref{appendix:density}). In this work, we focus on whether the search by QAOA can be made efficient by a simple yet robust pre-trianing scheme.

\subsection{QAOA Pre-training to Obtain Fixed Angles}

We take an approach to finding a good set of angles in the QAOA inspired by the search for parameters that has become synonymous with machine learning. Our simple yet robust pre-training algorithm is presented in alg. \ref{alg:pre-training}. 

The high-level intuition is as follows: train a collection of sets of angles, each on their own CVP instance drawn randomly from a training distribution, then evaluate how effective each set is at limiting the decay of the probability to sample the best solution on several random CVP instances drawn from a validation distribution. This is designed to find the set of angles best able to scale to larger problem instances, rather than is best at exploiting the nuances of a small training set.

\begin{algorithm}
    \caption{Pre-training for QAOA Angles}
    \label{alg:pre-training}
    
    \renewcommand{\algorithmicrequire}{\textbf{Input:}}
    \renewcommand{\algorithmicensure}{\textbf{Output:}}
    
    \begin{algorithmic}[1]
		\REQUIRE Training distribution $\mathcal{T}$ and set size $s_{t}$, validation distribution $\mathcal{V}$ and set size $s_{v}$, precision parameter $c$
		\ENSURE Optimal array of angles $\mathbf{a}_{op}$

        \STATE Initialise optimal array of angles $\mathbf{a}_{op}\leftarrow\mathbf{0}$
        \newline

        \FOR {some number of epochs}
            \STATE Initialise population of arrays of angles $A\leftarrow\{\}$
            \newline

            \FOR {$\_\leftarrow1,\dots,s_t$}
                \STATE Draw instance size $m\sim\mathcal{T}$
                \STATE Set $n\leftarrow m/\log m$
                \newline
                
                \STATE Construct the prime lattice $\mathcal{L}(B_{n,c})$
                \STATE Sample $m$-bit $N$ and define $\mathbf{t}$ accordingly
                \newline

                \STATE Initialise array of angles $\mathbf{a}\leftarrow\mathbf{a}_{op}$
                \STATE Optimise $\mathbf{a}$ for $\text{CVP}(\mathcal{L},\mathbf{t})$ by QAOA
                \STATE Update $A\leftarrow A\cup\{\mathbf{a}\}$
            \ENDFOR
            \newline

            \STATE Initialise best scaling $\alpha^*\leftarrow\infty$
            \newline

            \FOR {array $\mathbf{a}\in A$}
                \STATE Initialise set of data points $D\leftarrow\{\}$
                \newline

                \FOR {$\_\leftarrow1,\dots,s_v$}
                    \STATE Draw instance size $m\sim\mathcal{V}$
                    \STATE Set $n\leftarrow m/\log m$
                    \newline
                    
                    \STATE Construct the prime lattice $\mathcal{L}(B_{n,c})$
                    \STATE Sample $m$-bit $N$ and define $\mathbf{t}$ accordingly
                    \newline

                    \STATE Obtain set of vectors and probs $(\mathbf{v}_i,p_i)_{i=0,\dots,2^n}$ in the unit neighbourhood of $\mathbf{t}$ in $\mathcal{L}$
                    \STATE Initialise best distance $d^*\leftarrow\infty$ and prob $p^*\leftarrow0$
                    \newline
    
                    \FOR {$\mathbf{v},p\in(\mathbf{v}_i,p_i)_{i=0,\dots,2^n}$}
                        \STATE Compute distance $d\leftarrow\|\mathbf{t}-\mathbf{v}\|^2$
                        \STATE \textbf{if} $d<d^*$, update $d^*\leftarrow d$ and $p^*\leftarrow p$
                    \ENDFOR
                    \newline

                    \STATE Update $D\leftarrow D\cup\{(n,p^*)\}$
                \ENDFOR
                \newline

                \STATE Find $\alpha$ such that $p(n)=1/2^{\alpha n}$ is best-fit over $D$
                \STATE \textbf{if} $\alpha<\alpha^*$, update $\alpha^*\leftarrow\alpha$ and $\mathbf{a}_{op}\leftarrow\mathbf{a}$
            \ENDFOR
        \ENDFOR

    \end{algorithmic}
 
\end{algorithm}

\subsubsection{Notable design choices}

The issue of `overfitting' has been given careful consideration. A great source of attraction to this method is the ability to find a set of angles on small instances that may be solved relatively efficiently to save the expense in larger instances. However, finding a set of angles that is \textit{too} good for the smaller problems may not generalise well to larger problems, and so we must give great care to noticing when this is becoming the case in our search. This is where our method will greatly diverge from that of existing methods in \citet{brandao-2018} or \citet{boulebnane-2022}.

In each training instance, we initialise a set of angles from the best known angles at that time. We find that this works to make efficient the training loop where a head start can be offered. Our direct addressing of overfitting via a validation loop acts as mitigation for the potential that this `head start' may introduce.

We also make the explicit choice to work on the probability to sample \textit{the best} solution, rather than any solution improving on $\mathbf{b}_{op}$. This is done to: (1) avoid the unnecessary lattice reduction and computation of $\mathbf{b}_{op}$ in each validation instance; (2) ensures our angle sets tune the QAOA to find good solutions regardless of their numerosity (e.g. not biasing a set of parameters which are tuned on instances for which an atypical proportion of solutions are better than $\mathbf{b}_{op}$ by happenstance); and (3) avoid the complications that arise when $\mathbf{b}_{op}$ is already the best solution in the neighbourhood.

\subsection{Possible extensions and improvements}

This work presents a simple proof-of-concept for the use of pre-training in QAOA-based lattice cryptography. Future endeavours therefore have a plethora of extensions to be made to alg. \ref{alg:pre-training}, including but not limited to:
\begin{itemize}
    \item \textit{Cross-validation} to make for more efficient usage of data during an overfitting-aware scheme;
    \item \textit{Evaluating by cost} rather than by the probability to sample the minimum eigenstate, which may be more effective in larger neighbourhoods with greater variance of solution quality;
    \item \textit{Evolutionary algorithms} considering the population of arrays of angles $A$ as `evovling' over time, with the possibility for interaction, competition, and mutation between set of angles. 
\end{itemize}

\section{Experiments and Results} \label{sec:experiments}
\subsection{Pre-training and Angle Convergence}

\begin{figure}[h]
    \centering
    \textbf{Indicative Optimisation Landscape for QAOA}\par\medskip
    \includegraphics[width=\linewidth]{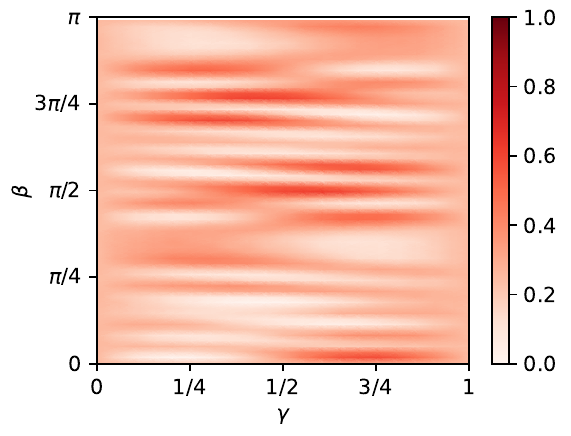}
    \caption{Plotting the the probability to successfully refine the solution (indicative of the underlying optimisation landscape) over a pair of parameters $(\gamma,\beta)$ in a $p=1$ layer QAOA for a random 3-qubit problem instance.}
    \label{fig:heatmap-3-qubits}
\end{figure}

\begin{figure}[h]
    \centering
    \textbf{Validation `Loss' Curves for Angle Pre-training}\par\medskip
    \includegraphics[width=\linewidth]{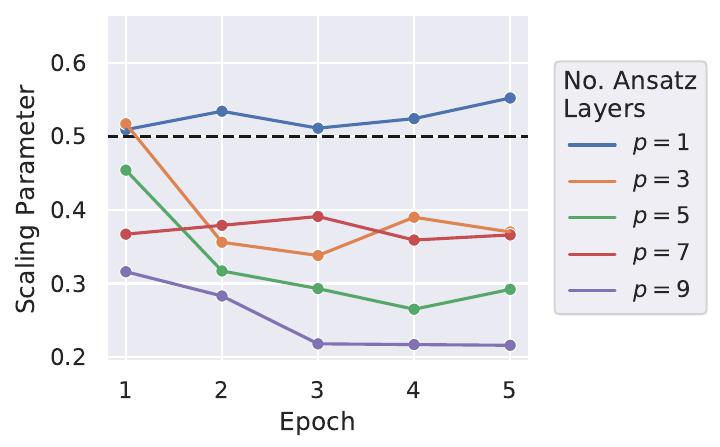}
    \caption{Validation performance during the pre-training of $p$-depth QAOA circuits. The `loss' being minimised is the scaling parameter $\alpha$ in the curve $q(n)=1/2^{\alpha n}$ over a dataset of validation points $(n_i,q_i)$. The dashed line denotes the speed-up offered by Grover's algorithm \cite{Grover-1996}.}
    \label{fig:-pretrain-val-performance}
\end{figure}

Ahead of our experimentation, we pre-train $p$-deep QAOA circuits according to alg. \ref{alg:pre-training} for $p=1,\dots,10$. The validation performances at each epoch are shown in fig. \ref{fig:-pretrain-val-performance}, showing general improvement for increasing $p$.

Discussion and prospective results are given for alternative training schemes in appendix \ref{appendix:alternative-training}.

\begin{figure}[h]
    \centering
    \textbf{Convergence of QAOA Angles by Bit-length}\par\medskip
    \includegraphics[width=.85\linewidth]{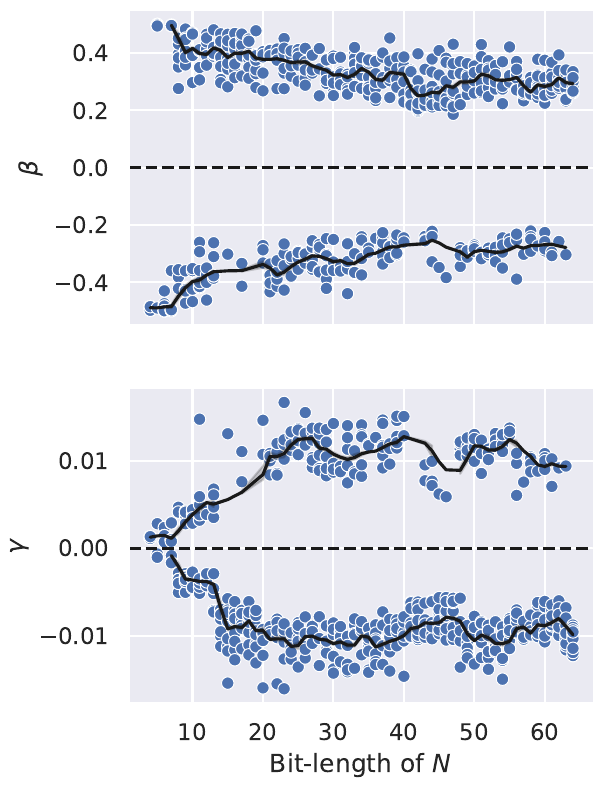}
    \caption{Convergence of the angles in a $p=1$ layer QAOA over different instances of varying bit-length.}
    \label{fig:qaoa-paramters-by-bit-length}
\end{figure}

Separately, we train a large number of $p=1$ layer QAOA circuits independently for random CVP instances and make note of the obtained values for $\beta$ and $\gamma$. We may plot these by the instance size, as shown in fig.  \ref{fig:qaoa-paramters-by-bit-length}, to observe the convergence of the angles for growing problem complexity. An indicative optimisation (for a small instance) landscape is exemplified by fig.  \ref{fig:heatmap-3-qubits}, though the landscape becomes increasingly flat as the problem complexity (here, lattice dimension) grows -- the dreaded ``barren plateau'' phenomenon \cite{Wang-2021, Uvarov-2021, Anschuetz-2022, Cerezo-2021b, Larocca-2024}.

From fig. \ref{fig:qaoa-paramters-by-bit-length}, we gain confidence that a stable set of angles can be easily learnt, and that they will have the capacity to scale to larger instances without the necessity to re-train nor fine-tune. 

Fig. \ref{fig:qaoa-paramters-by-bit-length} also highlights the issue of overfitting extremely well. Suppose we had followed a method similar to that of \citet{brandao-2018} and selected a single small instance on which to find our angles. The angles found on these smaller problems are not yet converged, unlike with later cases, and so are less likely to generalise well. Then any small instance we select will provide angles that are suboptimal in general. Hence, our scheme is advantageous in being aware of convergence and overfitting.

\subsection{Obtaining Statistics}

For our experiments, we conduct a broad numerical analysis to determine the trending relationship between refinement probability and lattice dimension, which then implies time-complexity via the expected number of queries to the QAOA circuit. Cases wherein the approximately found solution already represents the best solution in the unit neighbourhood are omitted.

First, generate a CVP for the $n$-dimensional lattice $\mathcal{L}(B_{n,c})$ for an $m$-bit composite integer $N$, where $n=m/\log m$ as specified by \citet{Schnorr-2021}. The method in \citet{Yan-2022} is only exemplified for three cases, however we simulate on the order of tens of thousands of cases so that an accurate scaling curve may be plotted to indicate asymptotic runtime.

We implement the QAOA solving a minimum-eigenstate problem for the Hamiltonian $\hat{H}_C$, as detailed in section \ref{sec:minimum-eigenstate-problem}, to obtain an outcome measurement $|\psi\rangle$ representing a binary string $\psi$ deciding whether to `step' in each of the reduced basis directions $D=[\mathbf{d}_1,\dots,\mathbf{d}_n]$ from $\mathbf{b}_{op}$. Hence, we may translate any $|\psi\rangle$ into the corresponding lattice vector $\mathbf{v}_{new}$ by performing $\mathbf{b}_{op}+\kappa\circ\psi D$, where $\circ$ denotes element-wise multiplication.

The probability to refine the solution (i.e. sample an improvement over $\mathbf{b}_{op}$) is ascertained by aggregating the probabilities for each $|\psi\rangle\mapsto\mathbf{v}_{new}$ for which $\|\mathbf{t}-\mathbf{v}_{new}\|^2<\|\mathbf{t}-\mathbf{b}_{op}\|^2$. This is the statistic whose decay we work to reduce with increasing lattice dimension $n$.

\subsection{Complexity Analysis for the Refinement}

Obtaining statistics for greatly many instances produces a dataset of points $(n_i,q_i)$, where $n_i=\log N_i/\log\log N_i$ is the `exact' lattice dimension computed directly from the composite integer $N_i$, and $q_i$ is the estimated probability to refine the approximation (obtained classically by the method in section \ref{sec:classical-approximation}). From $q_i$, we can expect to make $1/q_i$ queries to the circuit to yield the desired solution. 

These statistics are obtained by QAOA circuits of depths $p=1,\dots,10$ and plotted in fig. \ref{fig:performance-by-n}. In each case, we consider bit-lengths $4\leq m\leq128$, and thus lattice dimensions $3\leq n\leq18$. This is substantially larger than is considered in \citet{Yan-2022}, and should be large enough to reveal any scalability concerns \cite{Grebnev-2023, Khattar-2023, aboumrad-2023, Ducas-2021}.

Our optimal scaling is obtained, unsurprisingly, by $p=10$ layers, relating the refinement probability to lattice dimension as $q(n)\approx1/2^{0.225n}$, and thus indicating a time-complexity scaling as $O(2^{0.225n})$. This is more than a quadratic speed-up over the famous Grover's algorithm \cite{Grover-1996}, with far shallower depth and without requirement for fault tolerance. These findings mimic that of \citet{boulebnane-2022} for the improvement over Grover by fixed angles for QAOA. In fact, we find improvement with any depth $p>2$.

\begin{figure}[h]
    \centering
    \textbf{Relationship Between Scaling $\alpha$ and Depth $p$}\par\medskip
    \includegraphics[width=.8\linewidth]{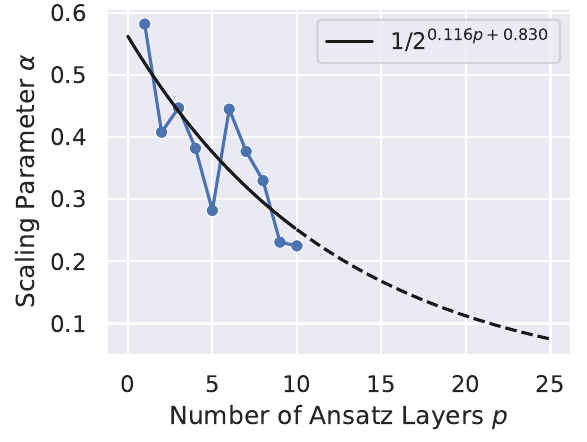}
    \caption{Extrapolating the relationship between the optimal scaling obtainable at each QAOA depth $p$, taken from fig. \ref{fig:performance-by-n}.}
    \label{fig:alpha-by-p}
\end{figure}

Our results indicate a promising relationship between the scaling parameter $\alpha$, which characterises the degree of the exponential decay by $1/2^{\alpha n}$, and the lattice dimension $n$. Fig. \ref{fig:alpha-by-p} fits an exponential curve to this relationship and extrapolates to greater depths. Optimistically, we estimate that the time complexity will reduce to $O(2^{0.1n})$ for $p>20$.

\subsection{Refinement Quality}

\begin{figure}[h]
    \centering
    \textbf{Quality of the Refinement by Lattice Dimension}\par\medskip
    \includegraphics[width=.8\linewidth]{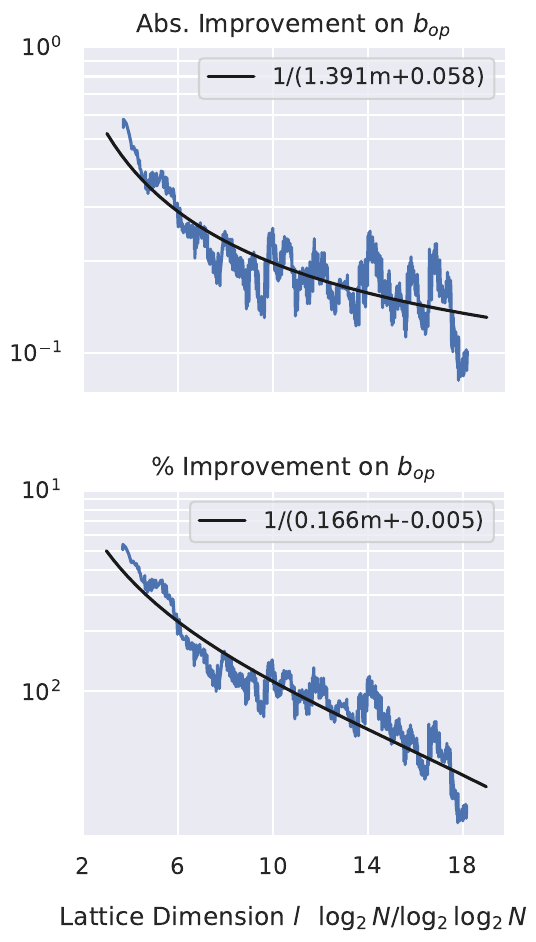}
    \caption{The improvement that the best solution in the neighbourhood obtains over $\mathbf{b}_{op}$ with increasing lattice dimension. A rolling average is given and a best-fit curve is shown.}
    \label{fig:quality-by-n}
\end{figure}

We will take brief notice of the scaling of the \textit{quality} of the refinement in fig. \ref{fig:quality-by-n}. While the improvement may appear small, we must note that \textit{any} improvement is significant due to the discrete nature of $\mathcal{L}$ -- an improvement indicates that an entirely different sr-pair has been identified, which comes with a different likelihood for usefulness in cryptography.

The quality of available solutions with this method is a primary concern in \citet{Grebnev-2023}, and is noted by other literature \cite{Khattar-2023, aboumrad-2023}. We consider data for which a refinement exists (i.e. situations in which the unit neighbourhood around $\mathbf{b}_{op}$ contains a better solution), and note a subexponential decay of quality in fig. \ref{fig:quality-by-n}. However, it is still unlikely that this quality is sufficient for sieving and thus factoring. 

We are optimistic that, given our findings for the potential quantum advantage in reducing the decay of refinement probability, the search space may be increased by dropping \citet{Schnorr-2021}'s contentious sublinear lattice dimension scheme (see appendix \ref{appendix:density}). 

\section{Conclusion}
In this work, we proposed a simple yet robust pre-training algorithm to use for fixed-angles QAOA. The method has been used on lattice-based problems, and enabled a heuristic analysis of the time-complexity of QAOA for sieving on the prime lattice \cite{Schnorr-1991, Schnorr-2013, Schnorr-2021}. 

Extending from the method in \citet{Yan-2022} for refining approximations to the closest vector problem as a reduction for the sieving problem, we indicate the possibility for a quantum advantage in searching for refinements via a fixed angle scheme for QAOA. In doing this, we hope to reveal the threat posed by newer variational approaches to lattice problems in cryptography.

Further work should explore whether this potential advantage persists as the search space for refinement grows -- say, for a neighbourhood encoding in $O(n\log n)$ qubits rather than $O(n)$. If the advantage is not lost in larger spaces, then lattice problems may be sufficiently challenged by these heuristic search methods, provided a set of angles can generalise well across instances. For a large enough neighbourhood (which is, in itself, not trivial to determine), there may be potential to find \textit{exact} solutions with a reasonable exponent in the time-complexity, though we expected that a highly constrained lattice should be assumed for such optimism. The prime lattice may be sufficiently constrained for explorations in this direction.

\subsection{Limitations}

Beyond the limitation due to insufficient search space, which has been discussed previously in connection with \citet{Yan-2022} and \citet{Schnorr-2021} regarding their ``sublinear factoring'' claim (see appendix \ref{appendix:density}), there are two limitations in this work that it is worth reiterating as a motivation for future research.

First, we only consider CVPs constructed on the prime lattice, as by their very nature (see section \ref{sec:method}), they are focused on sieving for sr-pairs. This gives our work a strong flavour of cryptanalysis, which serves as the main interest for finding a quantum advantage for lattice problems, though it may be argued to lack generality. Indeed, our success in finding a good set of angles may not be replicated with ease in more general CVPs, in which the structure has greater variance.

Second, this work \textit{simulates} quantum circuitry without considering the effect of noise. By working with relatively shallow-depth QAOA circuits, we remain optimistic that our findings can be experimentally ratified with real quantum hardware, perhaps with smaller advantage. We note that overly noisy hardware confounds the process of pre-training beyond practical use. We also note, however, that our algorithm can also be viewed as an early fault-tolerant one, since our approach could be applied to error-corrected circuits and would be better than, for example, naive Grover-type approaches, both in terms of time-complexity, and in terms of concrete number of logical qubits required (space-complexity).

\section*{Acknowledgements}

PW acknowledges support by EPSRC grants EP/T001062/1 and EP/Z53318X/1.

\bibliography{references}

\begin{thebibliography}{82}%
\makeatletter
\providecommand \@ifxundefined [1]{%
 \@ifx{#1\undefined}
}%
\providecommand \@ifnum [1]{%
 \ifnum #1\expandafter \@firstoftwo
 \else \expandafter \@secondoftwo
 \fi
}%
\providecommand \@ifx [1]{%
 \ifx #1\expandafter \@firstoftwo
 \else \expandafter \@secondoftwo
 \fi
}%
\providecommand \natexlab [1]{#1}%
\providecommand \enquote  [1]{``#1''}%
\providecommand \bibnamefont  [1]{#1}%
\providecommand \bibfnamefont [1]{#1}%
\providecommand \citenamefont [1]{#1}%
\providecommand \href@noop [0]{\@secondoftwo}%
\providecommand \href [0]{\begingroup \@sanitize@url \@href}%
\providecommand \@href[1]{\@@startlink{#1}\@@href}%
\providecommand \@@href[1]{\endgroup#1\@@endlink}%
\providecommand \@sanitize@url [0]{\catcode `\\12\catcode `\$12\catcode `\&12\catcode `\#12\catcode `\^12\catcode `\_12\catcode `\%12\relax}%
\providecommand \@@startlink[1]{}%
\providecommand \@@endlink[0]{}%
\providecommand \url  [0]{\begingroup\@sanitize@url \@url }%
\providecommand \@url [1]{\endgroup\@href {#1}{\urlprefix }}%
\providecommand \urlprefix  [0]{URL }%
\providecommand \Eprint [0]{\href }%
\providecommand \doibase [0]{https://doi.org/}%
\providecommand \selectlanguage [0]{\@gobble}%
\providecommand \bibinfo  [0]{\@secondoftwo}%
\providecommand \bibfield  [0]{\@secondoftwo}%
\providecommand \translation [1]{[#1]}%
\providecommand \BibitemOpen [0]{}%
\providecommand \bibitemStop [0]{}%
\providecommand \bibitemNoStop [0]{.\EOS\space}%
\providecommand \EOS [0]{\spacefactor3000\relax}%
\providecommand \BibitemShut  [1]{\csname bibitem#1\endcsname}%
\let\auto@bib@innerbib\@empty
\bibitem [{\citenamefont {Yan}\ \emph {et~al.}(2022)\citenamefont {Yan}, \citenamefont {Tan}, \citenamefont {Wei}, \citenamefont {Jiang}, \citenamefont {Wang}, \citenamefont {Wang}, \citenamefont {Luo}, \citenamefont {Duan}, \citenamefont {Liu}, \citenamefont {Shi}, \citenamefont {Fei}, \citenamefont {Meng}, \citenamefont {Han}, \citenamefont {Shan}, \citenamefont {Chen}, \citenamefont {Zhu}, \citenamefont {Zhang}, \citenamefont {Jin}, \citenamefont {Li}, \citenamefont {Song}, \citenamefont {Wang}, \citenamefont {Ma}, \citenamefont {Wang},\ and\ \citenamefont {Long}}]{Yan-2022}%
  \BibitemOpen
  \bibfield  {author} {\bibinfo {author} {\bibfnamefont {B.}~\bibnamefont {Yan}}, \bibinfo {author} {\bibfnamefont {Z.}~\bibnamefont {Tan}}, \bibinfo {author} {\bibfnamefont {S.}~\bibnamefont {Wei}}, \bibinfo {author} {\bibfnamefont {H.}~\bibnamefont {Jiang}}, \bibinfo {author} {\bibfnamefont {W.}~\bibnamefont {Wang}}, \bibinfo {author} {\bibfnamefont {H.}~\bibnamefont {Wang}}, \bibinfo {author} {\bibfnamefont {L.}~\bibnamefont {Luo}}, \bibinfo {author} {\bibfnamefont {Q.}~\bibnamefont {Duan}}, \bibinfo {author} {\bibfnamefont {Y.}~\bibnamefont {Liu}}, \bibinfo {author} {\bibfnamefont {W.}~\bibnamefont {Shi}}, \bibinfo {author} {\bibfnamefont {Y.}~\bibnamefont {Fei}}, \bibinfo {author} {\bibfnamefont {X.}~\bibnamefont {Meng}}, \bibinfo {author} {\bibfnamefont {Y.}~\bibnamefont {Han}}, \bibinfo {author} {\bibfnamefont {Z.}~\bibnamefont {Shan}}, \bibinfo {author} {\bibfnamefont {J.}~\bibnamefont {Chen}}, \bibinfo {author} {\bibfnamefont {X.}~\bibnamefont {Zhu}}, \bibinfo {author} {\bibfnamefont
  {C.}~\bibnamefont {Zhang}}, \bibinfo {author} {\bibfnamefont {F.}~\bibnamefont {Jin}}, \bibinfo {author} {\bibfnamefont {H.}~\bibnamefont {Li}}, \bibinfo {author} {\bibfnamefont {C.}~\bibnamefont {Song}}, \bibinfo {author} {\bibfnamefont {Z.}~\bibnamefont {Wang}}, \bibinfo {author} {\bibfnamefont {Z.}~\bibnamefont {Ma}}, \bibinfo {author} {\bibfnamefont {H.}~\bibnamefont {Wang}},\ and\ \bibinfo {author} {\bibfnamefont {G.-L.}\ \bibnamefont {Long}},\ }\href@noop {} {\bibinfo {title} {Factoring integers with sublinear resources on a superconducting quantum processor}} (\bibinfo {year} {2022}),\ \Eprint {https://arxiv.org/abs/2212.12372} {arXiv:2212.12372 [quant-ph]} \BibitemShut {NoStop}%
\bibitem [{\citenamefont {Rivest}\ \emph {et~al.}(1978)\citenamefont {Rivest}, \citenamefont {Shamir},\ and\ \citenamefont {Adleman}}]{Rivest-1978}%
  \BibitemOpen
  \bibfield  {author} {\bibinfo {author} {\bibfnamefont {R.~L.}\ \bibnamefont {Rivest}}, \bibinfo {author} {\bibfnamefont {A.}~\bibnamefont {Shamir}},\ and\ \bibinfo {author} {\bibfnamefont {L.}~\bibnamefont {Adleman}},\ }\bibfield  {title} {\bibinfo {title} {A method for obtaining digital signatures and public-key cryptosystems},\ }\href@noop {} {\bibfield  {journal} {\bibinfo  {journal} {Communications of the ACM}\ }\textbf {\bibinfo {volume} {21}},\ \bibinfo {pages} {120} (\bibinfo {year} {1978})}\BibitemShut {NoStop}%
\bibitem [{\citenamefont {Zhang}\ \emph {et~al.}(2024)\citenamefont {Zhang}, \citenamefont {Wang}, \citenamefont {Li},\ and\ \citenamefont {Wang}}]{Zhang-2024}%
  \BibitemOpen
  \bibfield  {author} {\bibinfo {author} {\bibfnamefont {D.}~\bibnamefont {Zhang}}, \bibinfo {author} {\bibfnamefont {H.}~\bibnamefont {Wang}}, \bibinfo {author} {\bibfnamefont {S.}~\bibnamefont {Li}},\ and\ \bibinfo {author} {\bibfnamefont {B.}~\bibnamefont {Wang}},\ }\bibfield  {title} {\bibinfo {title} {Progress in the prime factorization of large numbers},\ }\href@noop {} {\bibfield  {journal} {\bibinfo  {journal} {The Journal of Supercomputing}\ ,\ \bibinfo {pages} {1}} (\bibinfo {year} {2024})}\BibitemShut {NoStop}%
\bibitem [{\citenamefont {Shor}(1995)}]{Shor-1995}%
  \BibitemOpen
  \bibfield  {author} {\bibinfo {author} {\bibfnamefont {P.~W.}\ \bibnamefont {Shor}},\ }\bibfield  {title} {\bibinfo {title} {Polynomial-time algorithms for prime factorization and discrete logarithms on a quantum computer},\ }\href {https://api.semanticscholar.org/CorpusID:2337707} {\bibfield  {journal} {\bibinfo  {journal} {SIAM Rev.}\ }\textbf {\bibinfo {volume} {41}},\ \bibinfo {pages} {303} (\bibinfo {year} {1995})}\BibitemShut {NoStop}%
\bibitem [{\citenamefont {Lucero}\ \emph {et~al.}(2012)\citenamefont {Lucero}, \citenamefont {Barends}, \citenamefont {Chen}, \citenamefont {Kelly}, \citenamefont {Mariantoni}, \citenamefont {Megrant}, \citenamefont {O’Malley}, \citenamefont {Sank}, \citenamefont {Vainsencher}, \citenamefont {Wenner}, \citenamefont {White}, \citenamefont {Yin}, \citenamefont {Cleland},\ and\ \citenamefont {Martinis}}]{Lucero-2012}%
  \BibitemOpen
  \bibfield  {author} {\bibinfo {author} {\bibfnamefont {E.}~\bibnamefont {Lucero}}, \bibinfo {author} {\bibfnamefont {R.}~\bibnamefont {Barends}}, \bibinfo {author} {\bibfnamefont {Y.}~\bibnamefont {Chen}}, \bibinfo {author} {\bibfnamefont {J.}~\bibnamefont {Kelly}}, \bibinfo {author} {\bibfnamefont {M.}~\bibnamefont {Mariantoni}}, \bibinfo {author} {\bibfnamefont {A.}~\bibnamefont {Megrant}}, \bibinfo {author} {\bibfnamefont {P.}~\bibnamefont {O’Malley}}, \bibinfo {author} {\bibfnamefont {D.}~\bibnamefont {Sank}}, \bibinfo {author} {\bibfnamefont {A.}~\bibnamefont {Vainsencher}}, \bibinfo {author} {\bibfnamefont {J.}~\bibnamefont {Wenner}}, \bibinfo {author} {\bibfnamefont {T.}~\bibnamefont {White}}, \bibinfo {author} {\bibfnamefont {Y.}~\bibnamefont {Yin}}, \bibinfo {author} {\bibfnamefont {A.~N.}\ \bibnamefont {Cleland}},\ and\ \bibinfo {author} {\bibfnamefont {J.~M.}\ \bibnamefont {Martinis}},\ }\bibfield  {title} {\bibinfo {title} {Computing prime factors with a josephson phase qubit quantum
  processor},\ }\href {https://doi.org/10.1038/nphys2385} {\bibfield  {journal} {\bibinfo  {journal} {Nature Physics}\ }\textbf {\bibinfo {volume} {8}},\ \bibinfo {pages} {719–723} (\bibinfo {year} {2012})}\BibitemShut {NoStop}%
\bibitem [{\citenamefont {Lanyon}\ \emph {et~al.}(2007)\citenamefont {Lanyon}, \citenamefont {Weinhold}, \citenamefont {Langford}, \citenamefont {Barbieri}, \citenamefont {James}, \citenamefont {Gilchrist},\ and\ \citenamefont {White}}]{Lanyon-2007}%
  \BibitemOpen
  \bibfield  {author} {\bibinfo {author} {\bibfnamefont {B.~P.}\ \bibnamefont {Lanyon}}, \bibinfo {author} {\bibfnamefont {T.~J.}\ \bibnamefont {Weinhold}}, \bibinfo {author} {\bibfnamefont {N.~K.}\ \bibnamefont {Langford}}, \bibinfo {author} {\bibfnamefont {M.}~\bibnamefont {Barbieri}}, \bibinfo {author} {\bibfnamefont {D.~F.~V.}\ \bibnamefont {James}}, \bibinfo {author} {\bibfnamefont {A.}~\bibnamefont {Gilchrist}},\ and\ \bibinfo {author} {\bibfnamefont {A.~G.}\ \bibnamefont {White}},\ }\bibfield  {title} {\bibinfo {title} {Experimental demonstration of a compiled version of shor’s algorithm with quantum entanglement},\ }\bibfield  {journal} {\bibinfo  {journal} {Physical Review Letters}\ }\textbf {\bibinfo {volume} {99}},\ \href {https://doi.org/10.1103/physrevlett.99.250505} {10.1103/physrevlett.99.250505} (\bibinfo {year} {2007})\BibitemShut {NoStop}%
\bibitem [{\citenamefont {Lu}\ \emph {et~al.}(2007)\citenamefont {Lu}, \citenamefont {Browne}, \citenamefont {Yang},\ and\ \citenamefont {Pan}}]{Lu-2007}%
  \BibitemOpen
  \bibfield  {author} {\bibinfo {author} {\bibfnamefont {C.-Y.}\ \bibnamefont {Lu}}, \bibinfo {author} {\bibfnamefont {D.~E.}\ \bibnamefont {Browne}}, \bibinfo {author} {\bibfnamefont {T.}~\bibnamefont {Yang}},\ and\ \bibinfo {author} {\bibfnamefont {J.-W.}\ \bibnamefont {Pan}},\ }\bibfield  {title} {\bibinfo {title} {Demonstration of a compiled version of shor’s quantum factoring algorithm using photonic qubits},\ }\bibfield  {journal} {\bibinfo  {journal} {Physical Review Letters}\ }\textbf {\bibinfo {volume} {99}},\ \href {https://doi.org/10.1103/physrevlett.99.250504} {10.1103/physrevlett.99.250504} (\bibinfo {year} {2007})\BibitemShut {NoStop}%
\bibitem [{\citenamefont {Martín-López}\ \emph {et~al.}(2012)\citenamefont {Martín-López}, \citenamefont {Laing}, \citenamefont {Lawson}, \citenamefont {Alvarez}, \citenamefont {Zhou},\ and\ \citenamefont {O’Brien}}]{Martin-Lopez-2012}%
  \BibitemOpen
  \bibfield  {author} {\bibinfo {author} {\bibfnamefont {E.}~\bibnamefont {Martín-López}}, \bibinfo {author} {\bibfnamefont {A.}~\bibnamefont {Laing}}, \bibinfo {author} {\bibfnamefont {T.}~\bibnamefont {Lawson}}, \bibinfo {author} {\bibfnamefont {R.}~\bibnamefont {Alvarez}}, \bibinfo {author} {\bibfnamefont {X.-Q.}\ \bibnamefont {Zhou}},\ and\ \bibinfo {author} {\bibfnamefont {J.~L.}\ \bibnamefont {O’Brien}},\ }\bibfield  {title} {\bibinfo {title} {Experimental realization of shor’s quantum factoring algorithm using qubit recycling},\ }\href {https://doi.org/10.1038/nphoton.2012.259} {\bibfield  {journal} {\bibinfo  {journal} {Nature Photonics}\ }\textbf {\bibinfo {volume} {6}},\ \bibinfo {pages} {773–776} (\bibinfo {year} {2012})}\BibitemShut {NoStop}%
\bibitem [{\citenamefont {Bernstein}\ and\ \citenamefont {Lange}(2017)}]{Bernstein-2017}%
  \BibitemOpen
  \bibfield  {author} {\bibinfo {author} {\bibfnamefont {D.~J.}\ \bibnamefont {Bernstein}}\ and\ \bibinfo {author} {\bibfnamefont {T.}~\bibnamefont {Lange}},\ }\bibfield  {title} {\bibinfo {title} {Post-quantum cryptography},\ }\href@noop {} {\bibfield  {journal} {\bibinfo  {journal} {Nature}\ }\textbf {\bibinfo {volume} {549}},\ \bibinfo {pages} {188} (\bibinfo {year} {2017})}\BibitemShut {NoStop}%
\bibitem [{\citenamefont {Goldreich}\ \emph {et~al.}(1997)\citenamefont {Goldreich}, \citenamefont {Goldwasser},\ and\ \citenamefont {Halevi}}]{Goldreich-1997}%
  \BibitemOpen
  \bibfield  {author} {\bibinfo {author} {\bibfnamefont {O.}~\bibnamefont {Goldreich}}, \bibinfo {author} {\bibfnamefont {S.}~\bibnamefont {Goldwasser}},\ and\ \bibinfo {author} {\bibfnamefont {S.}~\bibnamefont {Halevi}},\ }\bibfield  {title} {\bibinfo {title} {Public-key cryptosystems from lattice reduction problems},\ }in\ \href@noop {} {\emph {\bibinfo {booktitle} {Advances in Cryptology—CRYPTO'97: 17th Annual International Cryptology Conference Santa Barbara, California, USA August 17--21, 1997 Proceedings 17}}}\ (\bibinfo {organization} {Springer},\ \bibinfo {year} {1997})\ pp.\ \bibinfo {pages} {112--131}\BibitemShut {NoStop}%
\bibitem [{\citenamefont {Hoffstein}\ \emph {et~al.}(1998)\citenamefont {Hoffstein}, \citenamefont {Pipher},\ and\ \citenamefont {Silverman}}]{Hoffstein-1998}%
  \BibitemOpen
  \bibfield  {author} {\bibinfo {author} {\bibfnamefont {J.}~\bibnamefont {Hoffstein}}, \bibinfo {author} {\bibfnamefont {J.}~\bibnamefont {Pipher}},\ and\ \bibinfo {author} {\bibfnamefont {J.~H.}\ \bibnamefont {Silverman}},\ }\bibfield  {title} {\bibinfo {title} {Ntru: A ring-based public key cryptosystem},\ }in\ \href@noop {} {\emph {\bibinfo {booktitle} {International algorithmic number theory symposium}}}\ (\bibinfo {organization} {Springer},\ \bibinfo {year} {1998})\ pp.\ \bibinfo {pages} {267--288}\BibitemShut {NoStop}%
\bibitem [{\citenamefont {Hoffstein}\ \emph {et~al.}(2001)\citenamefont {Hoffstein}, \citenamefont {Pipher},\ and\ \citenamefont {Silverman}}]{Hoffstein-2001}%
  \BibitemOpen
  \bibfield  {author} {\bibinfo {author} {\bibfnamefont {J.}~\bibnamefont {Hoffstein}}, \bibinfo {author} {\bibfnamefont {J.}~\bibnamefont {Pipher}},\ and\ \bibinfo {author} {\bibfnamefont {J.~H.}\ \bibnamefont {Silverman}},\ }\bibfield  {title} {\bibinfo {title} {Nss: An ntru lattice-based signature scheme},\ }in\ \href@noop {} {\emph {\bibinfo {booktitle} {Advances in Cryptology—EUROCRYPT 2001: International Conference on the Theory and Application of Cryptographic Techniques Innsbruck, Austria, May 6--10, 2001 Proceedings 20}}}\ (\bibinfo {organization} {Springer},\ \bibinfo {year} {2001})\ pp.\ \bibinfo {pages} {211--228}\BibitemShut {NoStop}%
\bibitem [{\citenamefont {Hoffstein}\ \emph {et~al.}(2003)\citenamefont {Hoffstein}, \citenamefont {Howgrave-Graham}, \citenamefont {Pipher}, \citenamefont {Silverman},\ and\ \citenamefont {Whyte}}]{Hoffstein-2003}%
  \BibitemOpen
  \bibfield  {author} {\bibinfo {author} {\bibfnamefont {J.}~\bibnamefont {Hoffstein}}, \bibinfo {author} {\bibfnamefont {N.}~\bibnamefont {Howgrave-Graham}}, \bibinfo {author} {\bibfnamefont {J.}~\bibnamefont {Pipher}}, \bibinfo {author} {\bibfnamefont {J.~H.}\ \bibnamefont {Silverman}},\ and\ \bibinfo {author} {\bibfnamefont {W.}~\bibnamefont {Whyte}},\ }\bibfield  {title} {\bibinfo {title} {Ntrusign: Digital signatures using the ntru lattice},\ }in\ \href@noop {} {\emph {\bibinfo {booktitle} {Cryptographers’ track at the RSA conference}}}\ (\bibinfo {organization} {Springer},\ \bibinfo {year} {2003})\ pp.\ \bibinfo {pages} {122--140}\BibitemShut {NoStop}%
\bibitem [{\citenamefont {Lyubashevsky}(2012)}]{Lyubashevsky-2012}%
  \BibitemOpen
  \bibfield  {author} {\bibinfo {author} {\bibfnamefont {V.}~\bibnamefont {Lyubashevsky}},\ }\bibfield  {title} {\bibinfo {title} {Lattice signatures without trapdoors},\ }in\ \href@noop {} {\emph {\bibinfo {booktitle} {Annual International Conference on the Theory and Applications of Cryptographic Techniques}}}\ (\bibinfo {organization} {Springer},\ \bibinfo {year} {2012})\ pp.\ \bibinfo {pages} {738--755}\BibitemShut {NoStop}%
\bibitem [{\citenamefont {Ducas}\ \emph {et~al.}(2013)\citenamefont {Ducas}, \citenamefont {Durmus}, \citenamefont {Lepoint},\ and\ \citenamefont {Lyubashevsky}}]{Ducas-2013}%
  \BibitemOpen
  \bibfield  {author} {\bibinfo {author} {\bibfnamefont {L.}~\bibnamefont {Ducas}}, \bibinfo {author} {\bibfnamefont {A.}~\bibnamefont {Durmus}}, \bibinfo {author} {\bibfnamefont {T.}~\bibnamefont {Lepoint}},\ and\ \bibinfo {author} {\bibfnamefont {V.}~\bibnamefont {Lyubashevsky}},\ }\bibfield  {title} {\bibinfo {title} {Lattice signatures and bimodal gaussians},\ }in\ \href@noop {} {\emph {\bibinfo {booktitle} {Annual Cryptology Conference}}}\ (\bibinfo {organization} {Springer},\ \bibinfo {year} {2013})\ pp.\ \bibinfo {pages} {40--56}\BibitemShut {NoStop}%
\bibitem [{\citenamefont {Bernstein}\ \emph {et~al.}(2018)\citenamefont {Bernstein}, \citenamefont {Chuengsatiansup}, \citenamefont {Lange},\ and\ \citenamefont {van Vredendaal}}]{Bernstein-2018}%
  \BibitemOpen
  \bibfield  {author} {\bibinfo {author} {\bibfnamefont {D.~J.}\ \bibnamefont {Bernstein}}, \bibinfo {author} {\bibfnamefont {C.}~\bibnamefont {Chuengsatiansup}}, \bibinfo {author} {\bibfnamefont {T.}~\bibnamefont {Lange}},\ and\ \bibinfo {author} {\bibfnamefont {C.}~\bibnamefont {van Vredendaal}},\ }\bibfield  {title} {\bibinfo {title} {Ntru prime: reducing attack surface at low cost},\ }in\ \href@noop {} {\emph {\bibinfo {booktitle} {Selected Areas in Cryptography--SAC 2017: 24th International Conference, Ottawa, ON, Canada, August 16-18, 2017, Revised Selected Papers 24}}}\ (\bibinfo {organization} {Springer},\ \bibinfo {year} {2018})\ pp.\ \bibinfo {pages} {235--260}\BibitemShut {NoStop}%
\bibitem [{\citenamefont {Coppersmith}\ and\ \citenamefont {Shamir}(1997)}]{Coppersmith-1997}%
  \BibitemOpen
  \bibfield  {author} {\bibinfo {author} {\bibfnamefont {D.}~\bibnamefont {Coppersmith}}\ and\ \bibinfo {author} {\bibfnamefont {A.}~\bibnamefont {Shamir}},\ }\bibfield  {title} {\bibinfo {title} {Lattice attacks on ntru},\ }in\ \href@noop {} {\emph {\bibinfo {booktitle} {International conference on the theory and applications of cryptographic techniques}}}\ (\bibinfo {organization} {Springer},\ \bibinfo {year} {1997})\ pp.\ \bibinfo {pages} {52--61}\BibitemShut {NoStop}%
\bibitem [{\citenamefont {Nguyen}\ and\ \citenamefont {Regev}(2006)}]{Nguyen-2006}%
  \BibitemOpen
  \bibfield  {author} {\bibinfo {author} {\bibfnamefont {P.~Q.}\ \bibnamefont {Nguyen}}\ and\ \bibinfo {author} {\bibfnamefont {O.}~\bibnamefont {Regev}},\ }\bibfield  {title} {\bibinfo {title} {Learning a parallelepiped: Cryptanalysis of ggh and ntru signatures},\ }in\ \href@noop {} {\emph {\bibinfo {booktitle} {Annual international conference on the theory and applications of cryptographic techniques}}}\ (\bibinfo {organization} {Springer},\ \bibinfo {year} {2006})\ pp.\ \bibinfo {pages} {271--288}\BibitemShut {NoStop}%
\bibitem [{\citenamefont {Ducas}\ and\ \citenamefont {Nguyen}(2012)}]{Ducas-2012}%
  \BibitemOpen
  \bibfield  {author} {\bibinfo {author} {\bibfnamefont {L.}~\bibnamefont {Ducas}}\ and\ \bibinfo {author} {\bibfnamefont {P.~Q.}\ \bibnamefont {Nguyen}},\ }\bibfield  {title} {\bibinfo {title} {Learning a zonotope and more: Cryptanalysis of ntrusign countermeasures},\ }in\ \href@noop {} {\emph {\bibinfo {booktitle} {International Conference on the Theory and Application of Cryptology and Information Security}}}\ (\bibinfo {organization} {Springer},\ \bibinfo {year} {2012})\ pp.\ \bibinfo {pages} {433--450}\BibitemShut {NoStop}%
\bibitem [{\citenamefont {Laarhoven}(2015)}]{Laarhoven-2015a}%
  \BibitemOpen
  \bibfield  {author} {\bibinfo {author} {\bibfnamefont {T.}~\bibnamefont {Laarhoven}},\ }\bibfield  {title} {\bibinfo {title} {Sieving for shortest vectors in lattices using angular locality-sensitive hashing},\ }in\ \href@noop {} {\emph {\bibinfo {booktitle} {Advances in Cryptology--CRYPTO 2015: 35th Annual Cryptology Conference, Santa Barbara, CA, USA, August 16-20, 2015, Proceedings, Part I 35}}}\ (\bibinfo {organization} {Springer},\ \bibinfo {year} {2015})\ pp.\ \bibinfo {pages} {3--22}\BibitemShut {NoStop}%
\bibitem [{\citenamefont {Laarhoven}\ and\ \citenamefont {de~Weger}(2015)}]{Laarhoven-2015b}%
  \BibitemOpen
  \bibfield  {author} {\bibinfo {author} {\bibfnamefont {T.}~\bibnamefont {Laarhoven}}\ and\ \bibinfo {author} {\bibfnamefont {B.}~\bibnamefont {de~Weger}},\ }\bibfield  {title} {\bibinfo {title} {Faster sieving for shortest lattice vectors using spherical locality-sensitive hashing},\ }in\ \href@noop {} {\emph {\bibinfo {booktitle} {Progress in Cryptology--LATINCRYPT 2015: 4th International Conference on Cryptology and Information Security in Latin America, Guadalajara, Mexico, August 23-26, 2015, Proceedings 4}}}\ (\bibinfo {organization} {Springer},\ \bibinfo {year} {2015})\ pp.\ \bibinfo {pages} {101--118}\BibitemShut {NoStop}%
\bibitem [{\citenamefont {Becker}\ \emph {et~al.}(2016)\citenamefont {Becker}, \citenamefont {Ducas}, \citenamefont {Gama},\ and\ \citenamefont {Laarhoven}}]{Becker-2016}%
  \BibitemOpen
  \bibfield  {author} {\bibinfo {author} {\bibfnamefont {A.}~\bibnamefont {Becker}}, \bibinfo {author} {\bibfnamefont {L.}~\bibnamefont {Ducas}}, \bibinfo {author} {\bibfnamefont {N.}~\bibnamefont {Gama}},\ and\ \bibinfo {author} {\bibfnamefont {T.}~\bibnamefont {Laarhoven}},\ }\bibfield  {title} {\bibinfo {title} {New directions in nearest neighbor searching with applications to lattice sieving},\ }in\ \href@noop {} {\emph {\bibinfo {booktitle} {Proceedings of the twenty-seventh annual ACM-SIAM symposium on Discrete algorithms}}}\ (\bibinfo {organization} {SIAM},\ \bibinfo {year} {2016})\ pp.\ \bibinfo {pages} {10--24}\BibitemShut {NoStop}%
\bibitem [{\citenamefont {Alagic}\ \emph {et~al.}(2022)\citenamefont {Alagic}, \citenamefont {Alagic}, \citenamefont {Apon}, \citenamefont {Cooper}, \citenamefont {Dang}, \citenamefont {Dang}, \citenamefont {Kelsey}, \citenamefont {Lichtinger}, \citenamefont {Liu}, \citenamefont {Miller} \emph {et~al.}}]{NIST-report-2022}%
  \BibitemOpen
  \bibfield  {author} {\bibinfo {author} {\bibfnamefont {G.}~\bibnamefont {Alagic}}, \bibinfo {author} {\bibfnamefont {G.}~\bibnamefont {Alagic}}, \bibinfo {author} {\bibfnamefont {D.}~\bibnamefont {Apon}}, \bibinfo {author} {\bibfnamefont {D.}~\bibnamefont {Cooper}}, \bibinfo {author} {\bibfnamefont {Q.}~\bibnamefont {Dang}}, \bibinfo {author} {\bibfnamefont {T.}~\bibnamefont {Dang}}, \bibinfo {author} {\bibfnamefont {J.}~\bibnamefont {Kelsey}}, \bibinfo {author} {\bibfnamefont {J.}~\bibnamefont {Lichtinger}}, \bibinfo {author} {\bibfnamefont {Y.-K.}\ \bibnamefont {Liu}}, \bibinfo {author} {\bibfnamefont {C.}~\bibnamefont {Miller}}, \emph {et~al.},\ }\bibfield  {title} {\bibinfo {title} {Status report on the third round of the nist post-quantum cryptography standardization process},\ }\href@noop {} {\bibfield  {journal} {\bibinfo  {journal} {CSRC}\ } (\bibinfo {year} {2022})}\BibitemShut {NoStop}%
\bibitem [{\citenamefont {Computer Security~Division}(2017)}]{NIST-website-2025}%
  \BibitemOpen
  \bibfield  {author} {\bibinfo {author} {\bibfnamefont {I.~T.~L.}\ \bibnamefont {Computer Security~Division}},\ }\href {https://csrc.nist.gov/projects/post-quantum-cryptography/post-quantum-cryptography-standardization} {\bibinfo {title} {Post-quantum cryptography standardization - post-quantum cryptography: Csrc}} (\bibinfo {year} {2017})\BibitemShut {NoStop}%
\bibitem [{\citenamefont {Pomerance}(1984)}]{Pomerance-1984}%
  \BibitemOpen
  \bibfield  {author} {\bibinfo {author} {\bibfnamefont {C.}~\bibnamefont {Pomerance}},\ }\bibfield  {title} {\bibinfo {title} {The quadratic sieve factoring algorithm},\ }in\ \href@noop {} {\emph {\bibinfo {booktitle} {Workshop on the Theory and Application of of Cryptographic Techniques}}}\ (\bibinfo {organization} {Springer},\ \bibinfo {year} {1984})\ pp.\ \bibinfo {pages} {169--182}\BibitemShut {NoStop}%
\bibitem [{\citenamefont {Davis}\ and\ \citenamefont {Holdridge}(1984)}]{Davis-1984}%
  \BibitemOpen
  \bibfield  {author} {\bibinfo {author} {\bibfnamefont {J.~A.}\ \bibnamefont {Davis}}\ and\ \bibinfo {author} {\bibfnamefont {D.~B.}\ \bibnamefont {Holdridge}},\ }\bibinfo {title} {Factorization using the quadratic sieve algorithm},\ in\ \href {https://doi.org/10.1007/978-1-4684-4730-9_9} {\emph {\bibinfo {booktitle} {Advances in Cryptology: Proceedings of Crypto 83}}},\ \bibinfo {editor} {edited by\ \bibinfo {editor} {\bibfnamefont {D.}~\bibnamefont {Chaum}}}\ (\bibinfo  {publisher} {Springer US},\ \bibinfo {address} {Boston, MA},\ \bibinfo {year} {1984})\ pp.\ \bibinfo {pages} {103--113}\BibitemShut {NoStop}%
\bibitem [{\citenamefont {Lenstra}\ and\ \citenamefont {Lenstra}(1993)}]{Lenstra-1993}%
  \BibitemOpen
  \bibfield  {author} {\bibinfo {author} {\bibfnamefont {A.~K.}\ \bibnamefont {Lenstra}}\ and\ \bibinfo {author} {\bibfnamefont {H.~W.}\ \bibnamefont {Lenstra}},\ }\href@noop {} {\emph {\bibinfo {title} {The development of the number field sieve}}},\ Vol.\ \bibinfo {volume} {1554}\ (\bibinfo  {publisher} {Springer Science \& Business Media},\ \bibinfo {year} {1993})\BibitemShut {NoStop}%
\bibitem [{\citenamefont {Briggs}(1998)}]{Briggs-1998}%
  \BibitemOpen
  \bibfield  {author} {\bibinfo {author} {\bibfnamefont {M.~E.}\ \bibnamefont {Briggs}},\ }\emph {\bibinfo {title} {An introduction to the general number field sieve}},\ \href@noop {} {Ph.D. thesis},\ \bibinfo  {school} {Virginia Tech} (\bibinfo {year} {1998})\BibitemShut {NoStop}%
\bibitem [{\citenamefont {Boudot}\ \emph {et~al.}(2022)\citenamefont {Boudot}, \citenamefont {Gaudry}, \citenamefont {Guillevic}, \citenamefont {Heninger}, \citenamefont {Thomé},\ and\ \citenamefont {Zimmermann}}]{boudot-2022}%
  \BibitemOpen
  \bibfield  {author} {\bibinfo {author} {\bibfnamefont {F.}~\bibnamefont {Boudot}}, \bibinfo {author} {\bibfnamefont {P.}~\bibnamefont {Gaudry}}, \bibinfo {author} {\bibfnamefont {A.}~\bibnamefont {Guillevic}}, \bibinfo {author} {\bibfnamefont {N.}~\bibnamefont {Heninger}}, \bibinfo {author} {\bibfnamefont {E.}~\bibnamefont {Thomé}},\ and\ \bibinfo {author} {\bibfnamefont {P.}~\bibnamefont {Zimmermann}},\ }\bibfield  {title} {\bibinfo {title} {The state of the art in integer factoring and breaking public-key cryptography},\ }\href {https://doi.org/10.1109/MSEC.2022.3141918} {\bibfield  {journal} {\bibinfo  {journal} {IEEE Security \& Privacy}\ }\textbf {\bibinfo {volume} {20}},\ \bibinfo {pages} {80} (\bibinfo {year} {2022})}\BibitemShut {NoStop}%
\bibitem [{\citenamefont {Schnorr}(1991)}]{Schnorr-1991}%
  \BibitemOpen
  \bibfield  {author} {\bibinfo {author} {\bibfnamefont {C.~P.}\ \bibnamefont {Schnorr}},\ }\bibfield  {title} {\bibinfo {title} {Factoring integers and computing discrete logarithms via diophantine approximation},\ }in\ \href@noop {} {\emph {\bibinfo {booktitle} {Advances in Cryptology --- EUROCRYPT '91}}},\ \bibinfo {editor} {edited by\ \bibinfo {editor} {\bibfnamefont {D.~W.}\ \bibnamefont {Davies}}}\ (\bibinfo  {publisher} {Springer Berlin Heidelberg},\ \bibinfo {address} {Berlin, Heidelberg},\ \bibinfo {year} {1991})\ pp.\ \bibinfo {pages} {281--293}\BibitemShut {NoStop}%
\bibitem [{\citenamefont {Schnorr}(2013)}]{Schnorr-2013}%
  \BibitemOpen
  \bibfield  {author} {\bibinfo {author} {\bibfnamefont {C.~P.}\ \bibnamefont {Schnorr}},\ }\bibfield  {title} {\bibinfo {title} {Factoring integers by cvp algorithms},\ }\href@noop {} {\bibfield  {journal} {\bibinfo  {journal} {Number Theory and Cryptography: Papers in Honor of Johannes Buchmann on the Occasion of His 60th Birthday}\ ,\ \bibinfo {pages} {73}} (\bibinfo {year} {2013})}\BibitemShut {NoStop}%
\bibitem [{\citenamefont {Schnorr}(2021)}]{Schnorr-2021}%
  \BibitemOpen
  \bibfield  {author} {\bibinfo {author} {\bibfnamefont {C.~P.}\ \bibnamefont {Schnorr}},\ }\href {https://eprint.iacr.org/2021/933} {\bibinfo {title} {Fast factoring integers by svp algorithms, corrected}},\ \bibinfo {howpublished} {Cryptology ePrint Archive, Paper 2021/933} (\bibinfo {year} {2021}),\ \bibinfo {note} {\url{https://eprint.iacr.org/2021/933}}\BibitemShut {NoStop}%
\bibitem [{\citenamefont {Farhi}\ \emph {et~al.}(2014)\citenamefont {Farhi}, \citenamefont {Goldstone},\ and\ \citenamefont {Gutmann}}]{Farhi-2014}%
  \BibitemOpen
  \bibfield  {author} {\bibinfo {author} {\bibfnamefont {E.}~\bibnamefont {Farhi}}, \bibinfo {author} {\bibfnamefont {J.}~\bibnamefont {Goldstone}},\ and\ \bibinfo {author} {\bibfnamefont {S.}~\bibnamefont {Gutmann}},\ }\bibfield  {title} {\bibinfo {title} {A quantum approximate optimization algorithm},\ }\href@noop {} {\bibfield  {journal} {\bibinfo  {journal} {arXiv preprint arXiv:1411.4028}\ } (\bibinfo {year} {2014})}\BibitemShut {NoStop}%
\bibitem [{\citenamefont {Grebnev}\ \emph {et~al.}(2023)\citenamefont {Grebnev}, \citenamefont {Gavreev}, \citenamefont {Kiktenko}, \citenamefont {Guglya}, \citenamefont {Efimov},\ and\ \citenamefont {Fedorov}}]{Grebnev-2023}%
  \BibitemOpen
  \bibfield  {author} {\bibinfo {author} {\bibfnamefont {S.~V.}\ \bibnamefont {Grebnev}}, \bibinfo {author} {\bibfnamefont {M.~A.}\ \bibnamefont {Gavreev}}, \bibinfo {author} {\bibfnamefont {E.~O.}\ \bibnamefont {Kiktenko}}, \bibinfo {author} {\bibfnamefont {A.~P.}\ \bibnamefont {Guglya}}, \bibinfo {author} {\bibfnamefont {A.~R.}\ \bibnamefont {Efimov}},\ and\ \bibinfo {author} {\bibfnamefont {A.~K.}\ \bibnamefont {Fedorov}},\ }\bibfield  {title} {\bibinfo {title} {Pitfalls of the sublinear qaoa-based factorization algorithm},\ }\href {https://doi.org/10.1109/access.2023.3336989} {\bibfield  {journal} {\bibinfo  {journal} {IEEE Access}\ }\textbf {\bibinfo {volume} {11}},\ \bibinfo {pages} {134760–134768} (\bibinfo {year} {2023})}\BibitemShut {NoStop}%
\bibitem [{\citenamefont {Aboumrad}\ \emph {et~al.}(2023)\citenamefont {Aboumrad}, \citenamefont {Widdows},\ and\ \citenamefont {Kaushik}}]{aboumrad-2023}%
  \BibitemOpen
  \bibfield  {author} {\bibinfo {author} {\bibfnamefont {W.}~\bibnamefont {Aboumrad}}, \bibinfo {author} {\bibfnamefont {D.}~\bibnamefont {Widdows}},\ and\ \bibinfo {author} {\bibfnamefont {A.}~\bibnamefont {Kaushik}},\ }\bibfield  {title} {\bibinfo {title} {Quantum and classical combinatorial optimizations applied to lattice-based factorization},\ }\href@noop {} {\bibfield  {journal} {\bibinfo  {journal} {arXiv preprint arXiv:2308.07804}\ } (\bibinfo {year} {2023})}\BibitemShut {NoStop}%
\bibitem [{\citenamefont {Khattar}\ and\ \citenamefont {Yosri}(2023)}]{Khattar-2023}%
  \BibitemOpen
  \bibfield  {author} {\bibinfo {author} {\bibfnamefont {T.}~\bibnamefont {Khattar}}\ and\ \bibinfo {author} {\bibfnamefont {N.}~\bibnamefont {Yosri}},\ }\bibfield  {title} {\bibinfo {title} {A comment on" factoring integers with sublinear resources on a superconducting quantum processor"},\ }\href@noop {} {\bibfield  {journal} {\bibinfo  {journal} {arXiv preprint arXiv:2307.09651}\ } (\bibinfo {year} {2023})}\BibitemShut {NoStop}%
\bibitem [{\citenamefont {Ducas}(2021)}]{Ducas-2021}%
  \BibitemOpen
  \bibfield  {author} {\bibinfo {author} {\bibfnamefont {L.}~\bibnamefont {Ducas}},\ }\href {https://github.com/lducas/SchnorrGate} {\bibinfo {title} {Lducas/schnorrgate: Testing schnorr’s factorization claim in sage}} (\bibinfo {year} {2021})\BibitemShut {NoStop}%
\bibitem [{\citenamefont {Vera}(2010)}]{Vera-2010}%
  \BibitemOpen
  \bibfield  {author} {\bibinfo {author} {\bibfnamefont {A.~I.}\ \bibnamefont {Vera}},\ }\bibfield  {title} {\bibinfo {title} {A note on integer factorization using lattices},\ }\href@noop {} {\bibfield  {journal} {\bibinfo  {journal} {arXiv preprint arXiv:1003.5461}\ } (\bibinfo {year} {2010})}\BibitemShut {NoStop}%
\bibitem [{\citenamefont {Boulebnane}\ and\ \citenamefont {Montanaro}(2022)}]{boulebnane-2022}%
  \BibitemOpen
  \bibfield  {author} {\bibinfo {author} {\bibfnamefont {S.}~\bibnamefont {Boulebnane}}\ and\ \bibinfo {author} {\bibfnamefont {A.}~\bibnamefont {Montanaro}},\ }\href@noop {} {\bibinfo {title} {Solving boolean satisfiability problems with the quantum approximate optimization algorithm}} (\bibinfo {year} {2022}),\ \Eprint {https://arxiv.org/abs/2208.06909} {arXiv:2208.06909 [quant-ph]} \BibitemShut {NoStop}%
\bibitem [{\citenamefont {Brandao}\ \emph {et~al.}(2018)\citenamefont {Brandao}, \citenamefont {Broughton}, \citenamefont {Farhi}, \citenamefont {Gutmann},\ and\ \citenamefont {Neven}}]{brandao-2018}%
  \BibitemOpen
  \bibfield  {author} {\bibinfo {author} {\bibfnamefont {F.~G. S.~L.}\ \bibnamefont {Brandao}}, \bibinfo {author} {\bibfnamefont {M.}~\bibnamefont {Broughton}}, \bibinfo {author} {\bibfnamefont {E.}~\bibnamefont {Farhi}}, \bibinfo {author} {\bibfnamefont {S.}~\bibnamefont {Gutmann}},\ and\ \bibinfo {author} {\bibfnamefont {H.}~\bibnamefont {Neven}},\ }\href@noop {} {\bibinfo {title} {For fixed control parameters the quantum approximate optimization algorithm's objective function value concentrates for typical instances}} (\bibinfo {year} {2018}),\ \Eprint {https://arxiv.org/abs/1812.04170} {arXiv:1812.04170 [quant-ph]} \BibitemShut {NoStop}%
\bibitem [{\citenamefont {Prokop}\ and\ \citenamefont {Wallden}(2025)}]{Prokop-2025}%
  \BibitemOpen
  \bibfield  {author} {\bibinfo {author} {\bibfnamefont {M.}~\bibnamefont {Prokop}}\ and\ \bibinfo {author} {\bibfnamefont {P.}~\bibnamefont {Wallden}},\ }\href {https://arxiv.org/abs/2502.05284} {\bibinfo {title} {Heuristic time complexity of nisq shortest-vector-problem solvers}} (\bibinfo {year} {2025}),\ \Eprint {https://arxiv.org/abs/2502.05284} {arXiv:2502.05284 [quant-ph]} \BibitemShut {NoStop}%
\bibitem [{\citenamefont {Priestley}(2025)}]{CVP-QAOA-code}%
  \BibitemOpen
  \bibfield  {author} {\bibinfo {author} {\bibfnamefont {B.}~\bibnamefont {Priestley}},\ }\href@noop {} {\bibinfo {title} {Code for the paper: A practical scalable approach to the cvp for sieving via qaoa with fixed angles}},\ \bibinfo {howpublished} {\url{https://github.com/BenPrie/qaoa-for-cvp}} (\bibinfo {year} {2025})\BibitemShut {NoStop}%
\bibitem [{Note1()}]{Note1}%
  \BibitemOpen
  \bibinfo {note} {We note that this does not conflict any known results on the asymptotic optimality of Grover, since QAOA is not a ``black-box'' oracle algorithm and uses the structure of the problem (via the problem Hamiltonian) in the way the ansatz is constructed.}\BibitemShut {Stop}%
\bibitem [{\citenamefont {Cerezo}\ \emph {et~al.}(2021{\natexlab{a}})\citenamefont {Cerezo}, \citenamefont {Arrasmith}, \citenamefont {Babbush}, \citenamefont {Benjamin}, \citenamefont {Endo}, \citenamefont {Fujii}, \citenamefont {McClean}, \citenamefont {Mitarai}, \citenamefont {Yuan}, \citenamefont {Cincio} \emph {et~al.}}]{Cerezo-2021}%
  \BibitemOpen
  \bibfield  {author} {\bibinfo {author} {\bibfnamefont {M.}~\bibnamefont {Cerezo}}, \bibinfo {author} {\bibfnamefont {A.}~\bibnamefont {Arrasmith}}, \bibinfo {author} {\bibfnamefont {R.}~\bibnamefont {Babbush}}, \bibinfo {author} {\bibfnamefont {S.~C.}\ \bibnamefont {Benjamin}}, \bibinfo {author} {\bibfnamefont {S.}~\bibnamefont {Endo}}, \bibinfo {author} {\bibfnamefont {K.}~\bibnamefont {Fujii}}, \bibinfo {author} {\bibfnamefont {J.~R.}\ \bibnamefont {McClean}}, \bibinfo {author} {\bibfnamefont {K.}~\bibnamefont {Mitarai}}, \bibinfo {author} {\bibfnamefont {X.}~\bibnamefont {Yuan}}, \bibinfo {author} {\bibfnamefont {L.}~\bibnamefont {Cincio}}, \emph {et~al.},\ }\bibfield  {title} {\bibinfo {title} {Variational quantum algorithms},\ }\href@noop {} {\bibfield  {journal} {\bibinfo  {journal} {Nature Reviews Physics}\ }\textbf {\bibinfo {volume} {3}},\ \bibinfo {pages} {625} (\bibinfo {year} {2021}{\natexlab{a}})}\BibitemShut {NoStop}%
\bibitem [{\citenamefont {Albrecht}\ \emph {et~al.}(2023)\citenamefont {Albrecht}, \citenamefont {Prokop}, \citenamefont {Shen},\ and\ \citenamefont {Wallden}}]{Albrecht-2023}%
  \BibitemOpen
  \bibfield  {author} {\bibinfo {author} {\bibfnamefont {M.~R.}\ \bibnamefont {Albrecht}}, \bibinfo {author} {\bibfnamefont {M.}~\bibnamefont {Prokop}}, \bibinfo {author} {\bibfnamefont {Y.}~\bibnamefont {Shen}},\ and\ \bibinfo {author} {\bibfnamefont {P.}~\bibnamefont {Wallden}},\ }\bibfield  {title} {\bibinfo {title} {Variational quantum solutions to the {S}hortest {V}ector {P}roblem},\ }\href {https://doi.org/10.22331/q-2023-03-02-933} {\bibfield  {journal} {\bibinfo  {journal} {{Quantum}}\ }\textbf {\bibinfo {volume} {7}},\ \bibinfo {pages} {933} (\bibinfo {year} {2023})}\BibitemShut {NoStop}%
\bibitem [{\citenamefont {Joseph}\ \emph {et~al.}(2021)\citenamefont {Joseph}, \citenamefont {Callison}, \citenamefont {Ling},\ and\ \citenamefont {Mintert}}]{Joseph-2021}%
  \BibitemOpen
  \bibfield  {author} {\bibinfo {author} {\bibfnamefont {D.}~\bibnamefont {Joseph}}, \bibinfo {author} {\bibfnamefont {A.}~\bibnamefont {Callison}}, \bibinfo {author} {\bibfnamefont {C.}~\bibnamefont {Ling}},\ and\ \bibinfo {author} {\bibfnamefont {F.}~\bibnamefont {Mintert}},\ }\bibfield  {title} {\bibinfo {title} {Two quantum ising algorithms for the shortest-vector problem},\ }\href {https://doi.org/10.1103/PhysRevA.103.032433} {\bibfield  {journal} {\bibinfo  {journal} {Phys. Rev. A}\ }\textbf {\bibinfo {volume} {103}},\ \bibinfo {pages} {032433} (\bibinfo {year} {2021})}\BibitemShut {NoStop}%
\bibitem [{\citenamefont {Babai}(1986)}]{Babai-1986}%
  \BibitemOpen
  \bibfield  {author} {\bibinfo {author} {\bibfnamefont {L.}~\bibnamefont {Babai}},\ }\bibfield  {title} {\bibinfo {title} {On lov{\'a}sz’lattice reduction and the nearest lattice point problem},\ }\href@noop {} {\bibfield  {journal} {\bibinfo  {journal} {Combinatorica}\ }\textbf {\bibinfo {volume} {6}},\ \bibinfo {pages} {1} (\bibinfo {year} {1986})}\BibitemShut {NoStop}%
\bibitem [{\citenamefont {Kraitchik}(1922)}]{kraitchik-1922}%
  \BibitemOpen
  \bibfield  {author} {\bibinfo {author} {\bibfnamefont {M.}~\bibnamefont {Kraitchik}},\ }\href@noop {} {\emph {\bibinfo {title} {Th{\'e}orie des nombres}}},\ Vol.~\bibinfo {volume} {1}\ (\bibinfo  {publisher} {Gauthier-Villars},\ \bibinfo {year} {1922})\BibitemShut {NoStop}%
\bibitem [{\citenamefont {Morrison}\ and\ \citenamefont {Brillhart}(1975)}]{Morrison-1975}%
  \BibitemOpen
  \bibfield  {author} {\bibinfo {author} {\bibfnamefont {M.~A.}\ \bibnamefont {Morrison}}\ and\ \bibinfo {author} {\bibfnamefont {J.}~\bibnamefont {Brillhart}},\ }\bibfield  {title} {\bibinfo {title} {A method of factoring and the factorization of f7},\ }\href@noop {} {\bibfield  {journal} {\bibinfo  {journal} {Mathematics of computation}\ }\textbf {\bibinfo {volume} {29}},\ \bibinfo {pages} {183} (\bibinfo {year} {1975})}\BibitemShut {NoStop}%
\bibitem [{\citenamefont {Dixon}(1981)}]{dixon-1981}%
  \BibitemOpen
  \bibfield  {author} {\bibinfo {author} {\bibfnamefont {J.~D.}\ \bibnamefont {Dixon}},\ }\bibfield  {title} {\bibinfo {title} {Asymptotically fast factorization of integers},\ }\href@noop {} {\bibfield  {journal} {\bibinfo  {journal} {Mathematics of computation}\ }\textbf {\bibinfo {volume} {36}},\ \bibinfo {pages} {255} (\bibinfo {year} {1981})}\BibitemShut {NoStop}%
\bibitem [{\citenamefont {Bennett}(2023)}]{Bennett-2023}%
  \BibitemOpen
  \bibfield  {author} {\bibinfo {author} {\bibfnamefont {H.}~\bibnamefont {Bennett}},\ }\bibfield  {title} {\bibinfo {title} {The complexity of the shortest vector problem},\ }\href {https://doi.org/10.1145/3586165.3586172} {\bibfield  {journal} {\bibinfo  {journal} {SIGACT News}\ }\textbf {\bibinfo {volume} {54}},\ \bibinfo {pages} {37–61} (\bibinfo {year} {2023})}\BibitemShut {NoStop}%
\bibitem [{\citenamefont {Regev}(2005)}]{Regev-2005}%
  \BibitemOpen
  \bibfield  {author} {\bibinfo {author} {\bibfnamefont {O.}~\bibnamefont {Regev}},\ }\bibfield  {title} {\bibinfo {title} {On lattices, learning with errors, random linear codes, and cryptography},\ }in\ \href {https://doi.org/10.1145/1060590.1060603} {\emph {\bibinfo {booktitle} {Proceedings of the Thirty-Seventh Annual ACM Symposium on Theory of Computing}}},\ \bibinfo {series and number} {STOC '05}\ (\bibinfo  {publisher} {Association for Computing Machinery},\ \bibinfo {address} {New York, NY, USA},\ \bibinfo {year} {2005})\ p.\ \bibinfo {pages} {84–93}\BibitemShut {NoStop}%
\bibitem [{\citenamefont {Bennett}\ \emph {et~al.}(2017)\citenamefont {Bennett}, \citenamefont {Golovnev},\ and\ \citenamefont {Stephens-Davidowitz}}]{bennett-2017}%
  \BibitemOpen
  \bibfield  {author} {\bibinfo {author} {\bibfnamefont {H.}~\bibnamefont {Bennett}}, \bibinfo {author} {\bibfnamefont {A.}~\bibnamefont {Golovnev}},\ and\ \bibinfo {author} {\bibfnamefont {N.}~\bibnamefont {Stephens-Davidowitz}},\ }\bibfield  {title} {\bibinfo {title} {On the quantitative hardness of cvp},\ }in\ \href@noop {} {\emph {\bibinfo {booktitle} {2017 IEEE 58th Annual Symposium on Foundations of Computer Science (FOCS)}}}\ (\bibinfo {organization} {IEEE},\ \bibinfo {year} {2017})\ pp.\ \bibinfo {pages} {13--24}\BibitemShut {NoStop}%
\bibitem [{\citenamefont {Micciancio}(2001)}]{micciancio-2001}%
  \BibitemOpen
  \bibfield  {author} {\bibinfo {author} {\bibfnamefont {D.}~\bibnamefont {Micciancio}},\ }\bibfield  {title} {\bibinfo {title} {The hardness of the closest vector problem with preprocessing},\ }\href@noop {} {\bibfield  {journal} {\bibinfo  {journal} {IEEE Transactions on Information Theory}\ }\textbf {\bibinfo {volume} {47}},\ \bibinfo {pages} {1212} (\bibinfo {year} {2001})}\BibitemShut {NoStop}%
\bibitem [{\citenamefont {Micciancio}\ and\ \citenamefont {Goldwasser}(2002)}]{micciancio-2002}%
  \BibitemOpen
  \bibfield  {author} {\bibinfo {author} {\bibfnamefont {D.}~\bibnamefont {Micciancio}}\ and\ \bibinfo {author} {\bibfnamefont {S.}~\bibnamefont {Goldwasser}},\ }\href@noop {} {\emph {\bibinfo {title} {Complexity of lattice problems: a cryptographic perspective}}},\ Vol.\ \bibinfo {volume} {671}\ (\bibinfo  {publisher} {Springer Science \& Business Media},\ \bibinfo {year} {2002})\BibitemShut {NoStop}%
\bibitem [{\citenamefont {Farhi}\ \emph {et~al.}(2000)\citenamefont {Farhi}, \citenamefont {Goldstone}, \citenamefont {Gutmann},\ and\ \citenamefont {Sipser}}]{Farhi-2000}%
  \BibitemOpen
  \bibfield  {author} {\bibinfo {author} {\bibfnamefont {E.}~\bibnamefont {Farhi}}, \bibinfo {author} {\bibfnamefont {J.}~\bibnamefont {Goldstone}}, \bibinfo {author} {\bibfnamefont {S.}~\bibnamefont {Gutmann}},\ and\ \bibinfo {author} {\bibfnamefont {M.}~\bibnamefont {Sipser}},\ }\bibfield  {title} {\bibinfo {title} {Quantum computation by adiabatic evolution},\ }\href@noop {} {\bibfield  {journal} {\bibinfo  {journal} {arXiv preprint quant-ph/0001106}\ } (\bibinfo {year} {2000})}\BibitemShut {NoStop}%
\bibitem [{\citenamefont {Farhi}\ \emph {et~al.}(2001)\citenamefont {Farhi}, \citenamefont {Goldstone}, \citenamefont {Gutmann}, \citenamefont {Lapan}, \citenamefont {Lundgren},\ and\ \citenamefont {Preda}}]{Farhi-2001}%
  \BibitemOpen
  \bibfield  {author} {\bibinfo {author} {\bibfnamefont {E.}~\bibnamefont {Farhi}}, \bibinfo {author} {\bibfnamefont {J.}~\bibnamefont {Goldstone}}, \bibinfo {author} {\bibfnamefont {S.}~\bibnamefont {Gutmann}}, \bibinfo {author} {\bibfnamefont {J.}~\bibnamefont {Lapan}}, \bibinfo {author} {\bibfnamefont {A.}~\bibnamefont {Lundgren}},\ and\ \bibinfo {author} {\bibfnamefont {D.}~\bibnamefont {Preda}},\ }\bibfield  {title} {\bibinfo {title} {A quantum adiabatic evolution algorithm applied to random instances of an np-complete problem},\ }\href@noop {} {\bibfield  {journal} {\bibinfo  {journal} {Science}\ }\textbf {\bibinfo {volume} {292}},\ \bibinfo {pages} {472} (\bibinfo {year} {2001})}\BibitemShut {NoStop}%
\bibitem [{\citenamefont {Zhou}\ \emph {et~al.}(2020)\citenamefont {Zhou}, \citenamefont {Wang}, \citenamefont {Choi}, \citenamefont {Pichler},\ and\ \citenamefont {Lukin}}]{Zhou-2020}%
  \BibitemOpen
  \bibfield  {author} {\bibinfo {author} {\bibfnamefont {L.}~\bibnamefont {Zhou}}, \bibinfo {author} {\bibfnamefont {S.-T.}\ \bibnamefont {Wang}}, \bibinfo {author} {\bibfnamefont {S.}~\bibnamefont {Choi}}, \bibinfo {author} {\bibfnamefont {H.}~\bibnamefont {Pichler}},\ and\ \bibinfo {author} {\bibfnamefont {M.~D.}\ \bibnamefont {Lukin}},\ }\bibfield  {title} {\bibinfo {title} {Quantum approximate optimization algorithm: Performance, mechanism, and implementation on near-term devices},\ }\bibfield  {journal} {\bibinfo  {journal} {Physical Review X}\ }\textbf {\bibinfo {volume} {10}},\ \href {https://doi.org/10.1103/physrevx.10.021067} {10.1103/physrevx.10.021067} (\bibinfo {year} {2020})\BibitemShut {NoStop}%
\bibitem [{\citenamefont {Bravyi}\ \emph {et~al.}(2019)\citenamefont {Bravyi}, \citenamefont {Browne}, \citenamefont {Calpin}, \citenamefont {Campbell}, \citenamefont {Gosset},\ and\ \citenamefont {Howard}}]{Bravyi-2019}%
  \BibitemOpen
  \bibfield  {author} {\bibinfo {author} {\bibfnamefont {S.}~\bibnamefont {Bravyi}}, \bibinfo {author} {\bibfnamefont {D.}~\bibnamefont {Browne}}, \bibinfo {author} {\bibfnamefont {P.}~\bibnamefont {Calpin}}, \bibinfo {author} {\bibfnamefont {E.}~\bibnamefont {Campbell}}, \bibinfo {author} {\bibfnamefont {D.}~\bibnamefont {Gosset}},\ and\ \bibinfo {author} {\bibfnamefont {M.}~\bibnamefont {Howard}},\ }\bibfield  {title} {\bibinfo {title} {Simulation of quantum circuits by low-rank stabilizer decompositions},\ }\href {https://doi.org/10.22331/q-2019-09-02-181} {\bibfield  {journal} {\bibinfo  {journal} {Quantum}\ }\textbf {\bibinfo {volume} {3}},\ \bibinfo {pages} {181} (\bibinfo {year} {2019})}\BibitemShut {NoStop}%
\bibitem [{\citenamefont {Grover}(1996)}]{Grover-1996}%
  \BibitemOpen
  \bibfield  {author} {\bibinfo {author} {\bibfnamefont {L.~K.}\ \bibnamefont {Grover}},\ }\href@noop {} {\bibinfo {title} {A fast quantum mechanical algorithm for database search}} (\bibinfo {year} {1996}),\ \Eprint {https://arxiv.org/abs/quant-ph/9605043} {arXiv:quant-ph/9605043 [quant-ph]} \BibitemShut {NoStop}%
\bibitem [{\citenamefont {Lenstra}\ \emph {et~al.}(1982)\citenamefont {Lenstra}, \citenamefont {Lenstra},\ and\ \citenamefont {Lov{\'a}sz}}]{Lenstra-1982}%
  \BibitemOpen
  \bibfield  {author} {\bibinfo {author} {\bibfnamefont {A.~K.}\ \bibnamefont {Lenstra}}, \bibinfo {author} {\bibfnamefont {H.~W.}\ \bibnamefont {Lenstra}},\ and\ \bibinfo {author} {\bibfnamefont {L.}~\bibnamefont {Lov{\'a}sz}},\ }\bibfield  {title} {\bibinfo {title} {Factoring polynomials with rational coefficients},\ }\href@noop {} {\bibfield  {journal} {\bibinfo  {journal} {Mathematische annalen}\ }\textbf {\bibinfo {volume} {261}},\ \bibinfo {pages} {515} (\bibinfo {year} {1982})}\BibitemShut {NoStop}%
\bibitem [{\citenamefont {Lucas}(2014)}]{lucas-2014}%
  \BibitemOpen
  \bibfield  {author} {\bibinfo {author} {\bibfnamefont {A.}~\bibnamefont {Lucas}},\ }\bibfield  {title} {\bibinfo {title} {Ising formulations of many np problems},\ }\href@noop {} {\bibfield  {journal} {\bibinfo  {journal} {Frontiers in physics}\ }\textbf {\bibinfo {volume} {2}},\ \bibinfo {pages} {5} (\bibinfo {year} {2014})}\BibitemShut {NoStop}%
\bibitem [{\citenamefont {Wang}\ \emph {et~al.}(2021)\citenamefont {Wang}, \citenamefont {Fontana}, \citenamefont {Cerezo}, \citenamefont {Sharma}, \citenamefont {Sone}, \citenamefont {Cincio},\ and\ \citenamefont {Coles}}]{Wang-2021}%
  \BibitemOpen
  \bibfield  {author} {\bibinfo {author} {\bibfnamefont {S.}~\bibnamefont {Wang}}, \bibinfo {author} {\bibfnamefont {E.}~\bibnamefont {Fontana}}, \bibinfo {author} {\bibfnamefont {M.}~\bibnamefont {Cerezo}}, \bibinfo {author} {\bibfnamefont {K.}~\bibnamefont {Sharma}}, \bibinfo {author} {\bibfnamefont {A.}~\bibnamefont {Sone}}, \bibinfo {author} {\bibfnamefont {L.}~\bibnamefont {Cincio}},\ and\ \bibinfo {author} {\bibfnamefont {P.~J.}\ \bibnamefont {Coles}},\ }\bibfield  {title} {\bibinfo {title} {Noise-induced barren plateaus in variational quantum algorithms},\ }\href@noop {} {\bibfield  {journal} {\bibinfo  {journal} {Nature communications}\ }\textbf {\bibinfo {volume} {12}},\ \bibinfo {pages} {6961} (\bibinfo {year} {2021})}\BibitemShut {NoStop}%
\bibitem [{\citenamefont {Uvarov}\ and\ \citenamefont {Biamonte}(2021)}]{Uvarov-2021}%
  \BibitemOpen
  \bibfield  {author} {\bibinfo {author} {\bibfnamefont {A.}~\bibnamefont {Uvarov}}\ and\ \bibinfo {author} {\bibfnamefont {J.~D.}\ \bibnamefont {Biamonte}},\ }\bibfield  {title} {\bibinfo {title} {On barren plateaus and cost function locality in variational quantum algorithms},\ }\href@noop {} {\bibfield  {journal} {\bibinfo  {journal} {Journal of Physics A: Mathematical and Theoretical}\ }\textbf {\bibinfo {volume} {54}},\ \bibinfo {pages} {245301} (\bibinfo {year} {2021})}\BibitemShut {NoStop}%
\bibitem [{\citenamefont {Anschuetz}\ and\ \citenamefont {Kiani}(2022)}]{Anschuetz-2022}%
  \BibitemOpen
  \bibfield  {author} {\bibinfo {author} {\bibfnamefont {E.~R.}\ \bibnamefont {Anschuetz}}\ and\ \bibinfo {author} {\bibfnamefont {B.~T.}\ \bibnamefont {Kiani}},\ }\bibfield  {title} {\bibinfo {title} {Quantum variational algorithms are swamped with traps},\ }\href@noop {} {\bibfield  {journal} {\bibinfo  {journal} {Nature Communications}\ }\textbf {\bibinfo {volume} {13}},\ \bibinfo {pages} {7760} (\bibinfo {year} {2022})}\BibitemShut {NoStop}%
\bibitem [{\citenamefont {Cerezo}\ \emph {et~al.}(2021{\natexlab{b}})\citenamefont {Cerezo}, \citenamefont {Sone}, \citenamefont {Volkoff}, \citenamefont {Cincio},\ and\ \citenamefont {Coles}}]{Cerezo-2021b}%
  \BibitemOpen
  \bibfield  {author} {\bibinfo {author} {\bibfnamefont {M.}~\bibnamefont {Cerezo}}, \bibinfo {author} {\bibfnamefont {A.}~\bibnamefont {Sone}}, \bibinfo {author} {\bibfnamefont {T.}~\bibnamefont {Volkoff}}, \bibinfo {author} {\bibfnamefont {L.}~\bibnamefont {Cincio}},\ and\ \bibinfo {author} {\bibfnamefont {P.~J.}\ \bibnamefont {Coles}},\ }\bibfield  {title} {\bibinfo {title} {Cost function dependent barren plateaus in shallow parametrized quantum circuits},\ }\href@noop {} {\bibfield  {journal} {\bibinfo  {journal} {Nature communications}\ }\textbf {\bibinfo {volume} {12}},\ \bibinfo {pages} {1791} (\bibinfo {year} {2021}{\natexlab{b}})}\BibitemShut {NoStop}%
\bibitem [{\citenamefont {Larocca}\ \emph {et~al.}(2024)\citenamefont {Larocca}, \citenamefont {Thanasilp}, \citenamefont {Wang}, \citenamefont {Sharma}, \citenamefont {Biamonte}, \citenamefont {Coles}, \citenamefont {Cincio}, \citenamefont {McClean}, \citenamefont {Holmes},\ and\ \citenamefont {Cerezo}}]{Larocca-2024}%
  \BibitemOpen
  \bibfield  {author} {\bibinfo {author} {\bibfnamefont {M.}~\bibnamefont {Larocca}}, \bibinfo {author} {\bibfnamefont {S.}~\bibnamefont {Thanasilp}}, \bibinfo {author} {\bibfnamefont {S.}~\bibnamefont {Wang}}, \bibinfo {author} {\bibfnamefont {K.}~\bibnamefont {Sharma}}, \bibinfo {author} {\bibfnamefont {J.}~\bibnamefont {Biamonte}}, \bibinfo {author} {\bibfnamefont {P.~J.}\ \bibnamefont {Coles}}, \bibinfo {author} {\bibfnamefont {L.}~\bibnamefont {Cincio}}, \bibinfo {author} {\bibfnamefont {J.~R.}\ \bibnamefont {McClean}}, \bibinfo {author} {\bibfnamefont {Z.}~\bibnamefont {Holmes}},\ and\ \bibinfo {author} {\bibfnamefont {M.}~\bibnamefont {Cerezo}},\ }\bibfield  {title} {\bibinfo {title} {A review of barren plateaus in variational quantum computing},\ }\href@noop {} {\bibfield  {journal} {\bibinfo  {journal} {arXiv preprint arXiv:2405.00781}\ } (\bibinfo {year} {2024})}\BibitemShut {NoStop}%
\bibitem [{\citenamefont {Joux}\ and\ \citenamefont {Stern}(1998)}]{Joux-1998}%
  \BibitemOpen
  \bibfield  {author} {\bibinfo {author} {\bibfnamefont {A.}~\bibnamefont {Joux}}\ and\ \bibinfo {author} {\bibfnamefont {J.}~\bibnamefont {Stern}},\ }\bibfield  {title} {\bibinfo {title} {Lattice reduction: A toolbox for the cryptanalyst},\ }\href@noop {} {\bibfield  {journal} {\bibinfo  {journal} {Journal of Cryptology}\ }\textbf {\bibinfo {volume} {11}},\ \bibinfo {pages} {161} (\bibinfo {year} {1998})}\BibitemShut {NoStop}%
\bibitem [{\citenamefont {Nguyen}\ and\ \citenamefont {Stern}(2000)}]{Nguyen-2000}%
  \BibitemOpen
  \bibfield  {author} {\bibinfo {author} {\bibfnamefont {P.~Q.}\ \bibnamefont {Nguyen}}\ and\ \bibinfo {author} {\bibfnamefont {J.}~\bibnamefont {Stern}},\ }\bibfield  {title} {\bibinfo {title} {Lattice reduction in cryptology: An update},\ }in\ \href@noop {} {\emph {\bibinfo {booktitle} {International Algorithmic Number Theory Symposium}}}\ (\bibinfo {organization} {Springer},\ \bibinfo {year} {2000})\ pp.\ \bibinfo {pages} {85--112}\BibitemShut {NoStop}%
\bibitem [{\citenamefont {Bremner}(2011)}]{bremner-2011}%
  \BibitemOpen
  \bibfield  {author} {\bibinfo {author} {\bibfnamefont {M.}~\bibnamefont {Bremner}},\ }\href@noop {} {\emph {\bibinfo {title} {Lattice basis reduction}}}\ (\bibinfo  {publisher} {CRC Press New York},\ \bibinfo {year} {2011})\BibitemShut {NoStop}%
\bibitem [{\citenamefont {W{\"u}bben}\ \emph {et~al.}(2011)\citenamefont {W{\"u}bben}, \citenamefont {Seethaler}, \citenamefont {Jalden},\ and\ \citenamefont {Matz}}]{Wubben-2011}%
  \BibitemOpen
  \bibfield  {author} {\bibinfo {author} {\bibfnamefont {D.}~\bibnamefont {W{\"u}bben}}, \bibinfo {author} {\bibfnamefont {D.}~\bibnamefont {Seethaler}}, \bibinfo {author} {\bibfnamefont {J.}~\bibnamefont {Jalden}},\ and\ \bibinfo {author} {\bibfnamefont {G.}~\bibnamefont {Matz}},\ }\bibfield  {title} {\bibinfo {title} {Lattice reduction},\ }\href@noop {} {\bibfield  {journal} {\bibinfo  {journal} {IEEE Signal Processing Magazine}\ }\textbf {\bibinfo {volume} {28}},\ \bibinfo {pages} {70} (\bibinfo {year} {2011})}\BibitemShut {NoStop}%
\bibitem [{\citenamefont {Thompson}(1996)}]{thompson-1996}%
  \BibitemOpen
  \bibfield  {author} {\bibinfo {author} {\bibfnamefont {A.~C.}\ \bibnamefont {Thompson}},\ }\href@noop {} {\emph {\bibinfo {title} {Minkowski geometry}}}\ (\bibinfo  {publisher} {Cambridge University Press},\ \bibinfo {year} {1996})\BibitemShut {NoStop}%
\bibitem [{\citenamefont {Ajtai}(1998)}]{ajtai-1998}%
  \BibitemOpen
  \bibfield  {author} {\bibinfo {author} {\bibfnamefont {M.}~\bibnamefont {Ajtai}},\ }\bibfield  {title} {\bibinfo {title} {The shortest vector problem in l2 is np-hard for randomized reductions},\ }in\ \href@noop {} {\emph {\bibinfo {booktitle} {Proceedings of the thirtieth annual ACM symposium on Theory of computing}}}\ (\bibinfo {year} {1998})\ pp.\ \bibinfo {pages} {10--19}\BibitemShut {NoStop}%
\bibitem [{\citenamefont {Ramaswami}(1949)}]{ramaswami-1949}%
  \BibitemOpen
  \bibfield  {author} {\bibinfo {author} {\bibfnamefont {V.}~\bibnamefont {Ramaswami}},\ }\bibfield  {title} {\bibinfo {title} {On the number of positive integers less than x and free of prime divisors greater than x\^{}c},\ }\href@noop {} {\bibfield  {journal} {\bibinfo  {journal} {Project Euclid}\ } (\bibinfo {year} {1949})}\BibitemShut {NoStop}%
\bibitem [{\citenamefont {de~Bruijn}(1951)}]{de-bruijn-1951}%
  \BibitemOpen
  \bibfield  {author} {\bibinfo {author} {\bibfnamefont {N.~G.}\ \bibnamefont {de~Bruijn}},\ }\bibfield  {title} {\bibinfo {title} {On the number of positive integers $\leq x $ and free of prime factors $> y$},\ }\href@noop {} {\bibfield  {journal} {\bibinfo  {journal} {Proceedings of the Koninklijke Nederlandse Akademie van Wetenschappen: Series A: Mathematical Sciences}\ }\textbf {\bibinfo {volume} {54}},\ \bibinfo {pages} {50} (\bibinfo {year} {1951})}\BibitemShut {NoStop}%
\bibitem [{\citenamefont {development team}(2023{\natexlab{a}})}]{fpylll}%
  \BibitemOpen
  \bibfield  {author} {\bibinfo {author} {\bibfnamefont {T.~F.}\ \bibnamefont {development team}},\ }\bibfield  {title} {\bibinfo {title} {{fpylll}, a {Python} wrapper for the {fplll} lattice reduction library, {Version}: 0.6.1}} (\bibinfo {year} {2023}{\natexlab{a}}),\ \bibinfo {note} {available at \url{https://github.com/fplll/fpylll}}\BibitemShut {NoStop}%
\bibitem [{\citenamefont {development team}(2023{\natexlab{b}})}]{fplll}%
  \BibitemOpen
  \bibfield  {author} {\bibinfo {author} {\bibfnamefont {T.~F.}\ \bibnamefont {development team}},\ }\bibfield  {title} {\bibinfo {title} {{fplll}, a lattice reduction library, {Version}: 5.4.5}} (\bibinfo {year} {2023}{\natexlab{b}}),\ \bibinfo {note} {available at \url{https://github.com/fplll/fplll}}\BibitemShut {NoStop}%
\bibitem [{\citenamefont {Harris}\ \emph {et~al.}(2020)\citenamefont {Harris}, \citenamefont {Millman}, \citenamefont {van~der Walt}, \citenamefont {Gommers}, \citenamefont {Virtanen}, \citenamefont {Cournapeau}, \citenamefont {Wieser}, \citenamefont {Taylor}, \citenamefont {Berg}, \citenamefont {Smith}, \citenamefont {Kern}, \citenamefont {Picus}, \citenamefont {Hoyer}, \citenamefont {van Kerkwijk}, \citenamefont {Brett}, \citenamefont {Haldane}, \citenamefont {del R{\'{i}}o}, \citenamefont {Wiebe}, \citenamefont {Peterson}, \citenamefont {G{\'{e}}rard-Marchant}, \citenamefont {Sheppard}, \citenamefont {Reddy}, \citenamefont {Weckesser}, \citenamefont {Abbasi}, \citenamefont {Gohlke},\ and\ \citenamefont {Oliphant}}]{numpy}%
  \BibitemOpen
  \bibfield  {author} {\bibinfo {author} {\bibfnamefont {C.~R.}\ \bibnamefont {Harris}}, \bibinfo {author} {\bibfnamefont {K.~J.}\ \bibnamefont {Millman}}, \bibinfo {author} {\bibfnamefont {S.~J.}\ \bibnamefont {van~der Walt}}, \bibinfo {author} {\bibfnamefont {R.}~\bibnamefont {Gommers}}, \bibinfo {author} {\bibfnamefont {P.}~\bibnamefont {Virtanen}}, \bibinfo {author} {\bibfnamefont {D.}~\bibnamefont {Cournapeau}}, \bibinfo {author} {\bibfnamefont {E.}~\bibnamefont {Wieser}}, \bibinfo {author} {\bibfnamefont {J.}~\bibnamefont {Taylor}}, \bibinfo {author} {\bibfnamefont {S.}~\bibnamefont {Berg}}, \bibinfo {author} {\bibfnamefont {N.~J.}\ \bibnamefont {Smith}}, \bibinfo {author} {\bibfnamefont {R.}~\bibnamefont {Kern}}, \bibinfo {author} {\bibfnamefont {M.}~\bibnamefont {Picus}}, \bibinfo {author} {\bibfnamefont {S.}~\bibnamefont {Hoyer}}, \bibinfo {author} {\bibfnamefont {M.~H.}\ \bibnamefont {van Kerkwijk}}, \bibinfo {author} {\bibfnamefont {M.}~\bibnamefont {Brett}}, \bibinfo {author} {\bibfnamefont
  {A.}~\bibnamefont {Haldane}}, \bibinfo {author} {\bibfnamefont {J.~F.}\ \bibnamefont {del R{\'{i}}o}}, \bibinfo {author} {\bibfnamefont {M.}~\bibnamefont {Wiebe}}, \bibinfo {author} {\bibfnamefont {P.}~\bibnamefont {Peterson}}, \bibinfo {author} {\bibfnamefont {P.}~\bibnamefont {G{\'{e}}rard-Marchant}}, \bibinfo {author} {\bibfnamefont {K.}~\bibnamefont {Sheppard}}, \bibinfo {author} {\bibfnamefont {T.}~\bibnamefont {Reddy}}, \bibinfo {author} {\bibfnamefont {W.}~\bibnamefont {Weckesser}}, \bibinfo {author} {\bibfnamefont {H.}~\bibnamefont {Abbasi}}, \bibinfo {author} {\bibfnamefont {C.}~\bibnamefont {Gohlke}},\ and\ \bibinfo {author} {\bibfnamefont {T.~E.}\ \bibnamefont {Oliphant}},\ }\bibfield  {title} {\bibinfo {title} {Array programming with {NumPy}},\ }\href {https://doi.org/10.1038/s41586-020-2649-2} {\bibfield  {journal} {\bibinfo  {journal} {Nature}\ }\textbf {\bibinfo {volume} {585}},\ \bibinfo {pages} {357} (\bibinfo {year} {2020})}\BibitemShut {NoStop}%
\bibitem [{\citenamefont {Developers}(2024)}]{cirq}%
  \BibitemOpen
  \bibfield  {author} {\bibinfo {author} {\bibfnamefont {C.}~\bibnamefont {Developers}},\ }\href {https://doi.org/10.5281/zenodo.11398048} {\bibinfo {title} {Cirq}} (\bibinfo {year} {2024})\BibitemShut {NoStop}%
\bibitem [{\citenamefont {team}\ and\ \citenamefont {collaborators}(2020)}]{qsim}%
  \BibitemOpen
  \bibfield  {author} {\bibinfo {author} {\bibfnamefont {Q.~A.}\ \bibnamefont {team}}\ and\ \bibinfo {author} {\bibnamefont {collaborators}},\ }\href {https://doi.org/10.5281/zenodo.4023103} {\bibinfo {title} {qsim}} (\bibinfo {year} {2020})\BibitemShut {NoStop}%
\bibitem [{\citenamefont {Gao}\ and\ \citenamefont {Han}(2012)}]{gao-2012}%
  \BibitemOpen
  \bibfield  {author} {\bibinfo {author} {\bibfnamefont {F.}~\bibnamefont {Gao}}\ and\ \bibinfo {author} {\bibfnamefont {L.}~\bibnamefont {Han}},\ }\bibfield  {title} {\bibinfo {title} {Implementing the nelder-mead simplex algorithm with adaptive parameters},\ }\href@noop {} {\bibfield  {journal} {\bibinfo  {journal} {Computational Optimization and Applications}\ }\textbf {\bibinfo {volume} {51}},\ \bibinfo {pages} {259} (\bibinfo {year} {2012})}\BibitemShut {NoStop}%
\bibitem [{\citenamefont {Virtanen}\ \emph {et~al.}(2020)\citenamefont {Virtanen}, \citenamefont {Gommers}, \citenamefont {Oliphant}, \citenamefont {Haberland}, \citenamefont {Reddy}, \citenamefont {Cournapeau}, \citenamefont {Burovski}, \citenamefont {Peterson}, \citenamefont {Weckesser}, \citenamefont {Bright}, \citenamefont {{van der Walt}}, \citenamefont {Brett}, \citenamefont {Wilson}, \citenamefont {Millman}, \citenamefont {Mayorov}, \citenamefont {Nelson}, \citenamefont {Jones}, \citenamefont {Kern}, \citenamefont {Larson}, \citenamefont {Carey}, \citenamefont {Polat}, \citenamefont {Feng}, \citenamefont {Moore}, \citenamefont {{VanderPlas}}, \citenamefont {Laxalde}, \citenamefont {Perktold}, \citenamefont {Cimrman}, \citenamefont {Henriksen}, \citenamefont {Quintero}, \citenamefont {Harris}, \citenamefont {Archibald}, \citenamefont {Ribeiro}, \citenamefont {Pedregosa}, \citenamefont {{van Mulbregt}},\ and\ \citenamefont {{SciPy 1.0 Contributors}}}]{scipy}%
  \BibitemOpen
  \bibfield  {author} {\bibinfo {author} {\bibfnamefont {P.}~\bibnamefont {Virtanen}}, \bibinfo {author} {\bibfnamefont {R.}~\bibnamefont {Gommers}}, \bibinfo {author} {\bibfnamefont {T.~E.}\ \bibnamefont {Oliphant}}, \bibinfo {author} {\bibfnamefont {M.}~\bibnamefont {Haberland}}, \bibinfo {author} {\bibfnamefont {T.}~\bibnamefont {Reddy}}, \bibinfo {author} {\bibfnamefont {D.}~\bibnamefont {Cournapeau}}, \bibinfo {author} {\bibfnamefont {E.}~\bibnamefont {Burovski}}, \bibinfo {author} {\bibfnamefont {P.}~\bibnamefont {Peterson}}, \bibinfo {author} {\bibfnamefont {W.}~\bibnamefont {Weckesser}}, \bibinfo {author} {\bibfnamefont {J.}~\bibnamefont {Bright}}, \bibinfo {author} {\bibfnamefont {S.~J.}\ \bibnamefont {{van der Walt}}}, \bibinfo {author} {\bibfnamefont {M.}~\bibnamefont {Brett}}, \bibinfo {author} {\bibfnamefont {J.}~\bibnamefont {Wilson}}, \bibinfo {author} {\bibfnamefont {K.~J.}\ \bibnamefont {Millman}}, \bibinfo {author} {\bibfnamefont {N.}~\bibnamefont {Mayorov}}, \bibinfo {author} {\bibfnamefont
  {A.~R.~J.}\ \bibnamefont {Nelson}}, \bibinfo {author} {\bibfnamefont {E.}~\bibnamefont {Jones}}, \bibinfo {author} {\bibfnamefont {R.}~\bibnamefont {Kern}}, \bibinfo {author} {\bibfnamefont {E.}~\bibnamefont {Larson}}, \bibinfo {author} {\bibfnamefont {C.~J.}\ \bibnamefont {Carey}}, \bibinfo {author} {\bibfnamefont {{\.I}.}~\bibnamefont {Polat}}, \bibinfo {author} {\bibfnamefont {Y.}~\bibnamefont {Feng}}, \bibinfo {author} {\bibfnamefont {E.~W.}\ \bibnamefont {Moore}}, \bibinfo {author} {\bibfnamefont {J.}~\bibnamefont {{VanderPlas}}}, \bibinfo {author} {\bibfnamefont {D.}~\bibnamefont {Laxalde}}, \bibinfo {author} {\bibfnamefont {J.}~\bibnamefont {Perktold}}, \bibinfo {author} {\bibfnamefont {R.}~\bibnamefont {Cimrman}}, \bibinfo {author} {\bibfnamefont {I.}~\bibnamefont {Henriksen}}, \bibinfo {author} {\bibfnamefont {E.~A.}\ \bibnamefont {Quintero}}, \bibinfo {author} {\bibfnamefont {C.~R.}\ \bibnamefont {Harris}}, \bibinfo {author} {\bibfnamefont {A.~M.}\ \bibnamefont {Archibald}}, \bibinfo {author}
  {\bibfnamefont {A.~H.}\ \bibnamefont {Ribeiro}}, \bibinfo {author} {\bibfnamefont {F.}~\bibnamefont {Pedregosa}}, \bibinfo {author} {\bibfnamefont {P.}~\bibnamefont {{van Mulbregt}}},\ and\ \bibinfo {author} {\bibnamefont {{SciPy 1.0 Contributors}}},\ }\bibfield  {title} {\bibinfo {title} {{{SciPy} 1.0: Fundamental Algorithms for Scientific Computing in Python}},\ }\href {https://doi.org/10.1038/s41592-019-0686-2} {\bibfield  {journal} {\bibinfo  {journal} {Nature Methods}\ }\textbf {\bibinfo {volume} {17}},\ \bibinfo {pages} {261} (\bibinfo {year} {2020})}\BibitemShut {NoStop}%
\end{thebibliography}%

\appendix 

\section{Lattice Reduction} \label{appendix:lattice-reduction}
The difficulty of any lattice problem is dictated in large part by the `quality' of the given basis $B$. A `good' basis is one consisting of short, relatively mutually orthogonal vectors, making navigation precise and intuitive. On the other hand, a `bad' basis consists of long, relatively mutually parallel vectors that confound the method of walking toward particular points by combinations of basis vectors. This intuition leads to public-key cryptosystems on lattices, for which a pair of good and bad bases imply private and public keys. 

Ideally then, we should like to start with a good basis, even if we are given a bad one to work from. The process of making a given basis `better' is referred to as \textit{lattice reduction}. Useful literature for developing an intuitive understanding for lattice reduction algorithms in cryptanalysis include \citet{Joux-1998}, \citet{Nguyen-2000}, and \citet{bremner-2011} in particular for an introduction. 

In this work, we consider the famous LLL-reduction algorithm due to \citet{Lenstra-1982}. For convenience, we give a brief description here. Useful texts include those aforementioned, or \citet{Wubben-2011}. Discussion here draws also from \citet{Schnorr-2021}.

\begin{definition} [QR-decomposition]
    Any basis matrix $B$ has the unique decomposition $B=QR$, where $Q\in\mathbb{R}^{n\times m}$ is isometric (with pairwise orthogonal column vectors of unit length) and $R=[r_{i,j}]_{1\leq i,j\leq m}\in\mathbb{R}^{m\times m}$ is upper triangular with positive diagonal entries $r_{i,i}$.
\end{definition}

Furthermore, $R=\text{GNF}(B)$ is the \textit{generic normal form} of $B$, whose Gram-Schmidt coefficients $\mu_{j,i}=r_{i,j}/r_{i,i}$ are rational for integer matrices.

\begin{definition} [LLL Reduction]
    A basis $B$ is $\delta$-LLL reduced, if $|\mu_{i,j}|$ for all $i<j$, and $\delta r^2_{i,i}\leq r^2_{i,i+1}+r^2_{i+1,i+1}$ for $i=1,\dots,n-1$. 
\end{definition}

We enforce that $\frac{1}{4}<\delta\leq1$. \citet{Lenstra-1982} show that any basis can be $\delta$-LLL reduced for $\delta<1$ in polynomial time, and that they approximate the successive minima well.

\section{Sublinearity of Schnorr's Method} \label{appendix:density}
This section provides supplementary detail for the lattice dimension scheme presented by Schnorr \cite{Schnorr-1991, Schnorr-2013, Schnorr-2021} that gives rise to the sublinearity claim later championed in \citet{Schnorr-2021} and \citet{Yan-2022} to ``destroy the RSA cryptosystem''. We then discuss the evidence against the validity of this scheme \cite{Grebnev-2023, Khattar-2023, aboumrad-2023, Ducas-2021}, with particular focus on the findings of \citet{aboumrad-2023}.

\subsection{The Sublinear Scheme}

By Minkowski's first theorem (see \cite{thompson-1996}), the shortest vector $\lambda_1$ for any full rank $n$-dimensional lattice $\mathcal{L}\subseteq\mathbb{R}^n$ is bounded from above as
\begin{equation}
    \lambda_1(\mathcal{L})^2\leq n\cdot(\det\mathcal{L})^{2/n}\ .
\end{equation}

The discrepancy between the \textit{real} shortest vector $\lambda_1$ and this bound can be measured by the \textit{relative density} $\text{rd}(\mathcal{L})$ of the lattice, which gives a ratio between $\lambda_1$ and the upper bound estimated by the Hermite constant;
\begin{equation}
    \text{rd}(\mathcal{L}):=\frac{\lambda_1(\mathcal{L})}{\sqrt{\gamma_n}(\det\mathcal{L})^{1/n}}\ ,
\end{equation}
where $\gamma_n$ is the Hermite constant of definition \ref{def:hermite}.

\citet{Schnorr-1991} made the critical assumption that a random lattice $\mathcal{L}$ with size-ordered basis $B=[\mathbf{b}_1,\dots,\mathbf{b}_m]$ has a relative density satisfying
\begin{equation}
    \text{rd}(\mathcal{L})\leq\bigg(\sqrt{\frac{e\pi}{2n}}\cdot\frac{\lambda_1(\mathcal{L})}{\|\mathbf{b}_1\|}\bigg)^{1/2}\ ,
\end{equation}
and since $\lambda_1(\mathcal{L})\|/\mathbf{b_1}\|\leq1$, this then leads to
\begin{equation} \label{eq:relative-density-assumption}
    \text{rd}(\mathcal{L})=\frac{\lambda_1(\mathcal{L})}{\sqrt{\gamma_n}(\det\mathcal{L})^{1/n}}\leq\bigg(\frac{e\pi}{2n}\bigg)^{1/4}\ .
\end{equation}

We can then propose that, for the lattice $\mathcal{L}(B_{n,c})$ whose dimension satisfies $n=2c\log N/\log\log N$, and whose relative density satisfies eq. (\ref{eq:relative-density-assumption}), there exists some vector $\mathbf{v}\in\mathcal{L}(B_{n,c})$ such that $\|\mathbf{v}-\mathbf{t}\|^2=O(\log N)$ for any target vector $\mathbf{t}\in\text{Span}(B_{n,c})$. Proofs for this proposition are given in \citet{Schnorr-2021} and in the appendix of \citet{Yan-2022}.

\subsection{Validity of the Scheme and Assumption}

Firstly, the sublinear scheme is only of practical use if the reduction from factoring to the search for sr-pairs works in polynomial time, given the heuristic assumptions of density culminating in eq. (\ref{eq:relative-density-assumption}). Practical implementations, such as \citet{Ducas-2021}, dispute that this assumption successfully scales well enough to be cryptographically significant. Theoretical considerations tend to agree with this disputatious attitude \cite{Grebnev-2023}.

Secondly, since the foundations of \citet{Schnorr-1991}'s guarantees were formulated in the early 1990s, the theory of lattice problems has developed tremendously, including the reliance of most popular CVP heuristics on $\ell_2$-measurements (e.g. \cite{ajtai-1998}). This leaves the proofs in \citet{Schnorr-1991} (in particular, Lemma 2) with limited practical bearing \cite{aboumrad-2023}.

Thirdly, \citet{aboumrad-2023} use theoretic results \cite{ramaswami-1949, de-bruijn-1951} to estimate the density of $p_N$-smooth numbers as $N$ grows (note that $p_N$-smooth here refers to integers whose prime factors are not greater than $N$). Combining Dickman's function with the prime number theorem allows them to estimate that the proportion of $p_N$-smooth numbers below $N$ that may be collected under the sublinear lattice dimension scheme is exponentially small (see section 3.1 of \citet{aboumrad-2023}). 

For small $N$, the assumption is acceptable, but the exponential decay of available smooth numbers renders the assumption unacceptable for large $N$, and hence the scalability of Schnorr's method \cite{Schnorr-1991, Schnorr-2013, Schnorr-2021} should be doubted under a sublinear scheme. This decay is the primary cause for the necessity for exponentially many CVP instances needing to be solved for a factorisation to be possible -- we need exponentially many solutions to be sure of finding enough unique sr-pairs for post-processing.

\begin{figure*}
    \centering
    \textbf{Using a Single CVP Instance to Find and Fix Angles}\par\medskip
    \includegraphics[width=.8\textwidth]{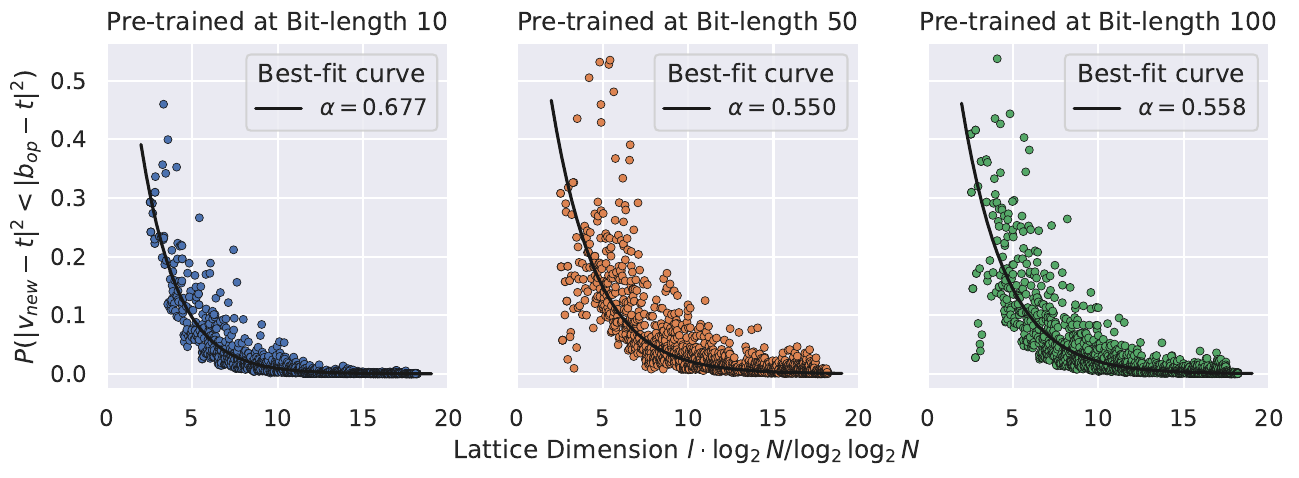}
    \caption{Probability to refine the approximate solution by lattice dimension for three distinct QAOA circuits with angles specified by a single (random) instance. The size of these instance (given in bit-length of $N$) is indicated above each plot.}
    \label{fig:alternative-random-instance}
\end{figure*}

\begin{figure*}
    \centering
    \textbf{Optimising Each CVP Instance Independently (No Pretraining)}\par\medskip
    \includegraphics[width=.8\textwidth]{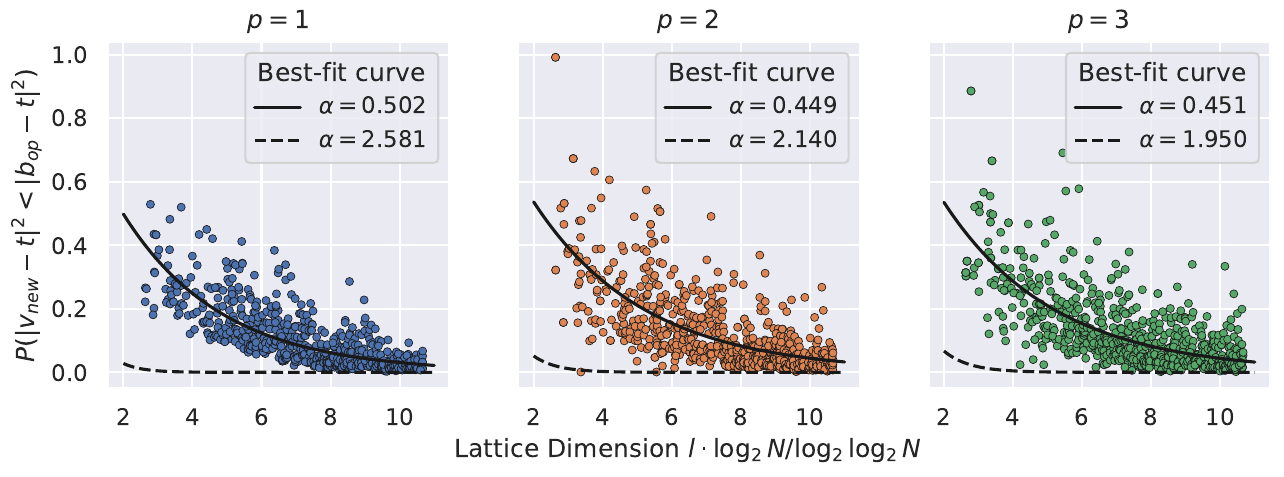}
    \caption{Similarly to fig. \ref{fig:alternative-random-instance}, though without any pre-training scheme. That is, each instance of the QAOA obtains its angles independently. We show here also a dashed line representing the scaling when also considering cases in which no refinement exists within the neighbourhood.}
    \label{fig:alternative-none}
\end{figure*}

\section{Alternative Angle Selection Schemes} \label{appendix:alternative-training}
This appendix presents some prospective results for our scaling analysis with alternative methods for angle selection. This serves to contextualise the success of our proposed pre-training scheme.

\subsection{Random Instance Pre-training}

The most straightforward way to obtain angles -- short of drawing them randomly -- is to train on a single (random) instance and take the consequent angles. This is the impression left by \citet{brandao-2018}.

Ideally, the single instance we choose as our pre-training instance is small enough that the angles may be obtained relatively efficiently, but large enough that the resultant angles are general (see fig. \ref{fig:qaoa-paramters-by-bit-length} to get a sense for the convergence of the angles by instance size).

Fig. \ref{fig:alternative-random-instance} gives the time complexity (mimicking the subplots in fig. \ref{fig:performance-by-n}) one might expect to obtain in general having pre-trained by a single random instance at the indicated size. This is roughly inline with our expectation that bit-lengths $<20$ are ineffective for obtaining generalisable angles. We further notice that increasing bit-length does not appear to continue reducing the probability decay.

Our preliminary conclusion is that pre-training by a single random instance has the limit of Grover's speed up ($\alpha\approx\frac{1}{2}$). More instances are required to gather the information necessary to find general angles, as in our proposed scheme in section \ref{sec:method}.

\subsection{No Pre-training}

We would also like to show propsective results from a lack of pre-training altogether. Here, we obtain the angles in each instance independently, as is currently standard in QAOA. 

This, of course, results in immense computational effort when considering greatly many instances as we do, hence we are unable to explore to the largest lattice dimensions. Already, this is a decisive difference between a fixed angle scheme and the absence thereof -- when you are happy to omit the optimisation of angles, we can set out immediately querying the circuit!

The prospective results, shown in fig. \ref{fig:alternative-none}, are better than for random instance pre-training, though have a similar apprehension for continued improvement. The improvement over random instance pre-training is expected (after all, each instance finds its own `perfect' angles), but an overall lack of improvement over our own pre-training scheme (see fig. \ref{fig:performance-by-n}) is surprising. Indeed, exploring to greater depths may be the deciding factor in demonstrating an ability to withstand the exponential decay of probability.

Again, we conclude that pre-training is an indispensable tool in the usage of QAOA, and leads to the potential for a quantum advantage that we highlight in this work.

\section{Implementation Details} \label{appendix:implementation}
Our work was produced in Python (version 3.8.18), and is available at \cite{CVP-QAOA-code}. 

LLL-reduction \cite{Lenstra-1982} and Babai's nearest plane algorithm \cite{Babai-1986} are implemented via FPyLLL (version 0.6.1; \cite{fpylll}), a Python interface for the lattice algorithm library FPLLL (version 5.4.5; \cite{fplll}). This further relied on NumPy (version 1.22.3; \cite{numpy}).

For the quantum portion of our implementation, we use Google Cirq (version 1.1.0; \cite{cirq}) and the accompanying qsimcirc (version 0.18.0; \cite{qsim}) for simulation. Unfortunately, Cirq does not provide an implementation of QAOA (yet), so we have derived our own implementation from \citet{Khattar-2023}.

For the results presented in this work, we use the Nelder-Mead simplex algorithm \cite{gao-2012} for minimisation of the expectation values of the observables in our circuit via SciPy (version 1.10.1; \cite{scipy}).

\end{document}